\documentclass[a4paper, 12pt]{article}
\usepackage{epsfig,amssymb,euscript,xspace, color}
\usepackage{amsmath,jheppub,mathtools,empheq,amsthm}
\usepackage{graphicx}
\usepackage{verbatim}
\usepackage{cancel}

\usepackage[T1]{fontenc} 
\usepackage{tikz,caption,subcaption,marvosym} 
\usetikzlibrary{decorations.markings,arrows,snakes}
\usepackage{epsfig,amssymb,euscript,xspace, color}
\usepackage{amsmath,mathtools,rotating}

\def\bea{\begin{eqnarray}}
\def\eea{\end{eqnarray}}
\def\be{\begin{equation}}
\def\ee{\end{equation}}



\begin{document}

\title{
A Chiral $\Lambda$-$\mathfrak{bms}_4$ Symmetry of AdS$_4$ Gravity }
\author{Nishant Gupta and Nemani V. Suryanarayana}
\affiliation{Institute of Mathematical Sciences,\\ Taramani, Chennai 600113, India \\
\& \\ Homi Bhabha National Institute,  \\ Anushakti Nagar}
\emailAdd{nishantg, nemani@imsc.res.in}

\abstract{Generalising the chiral boundary conditions of $\mathbb{R}^{1,3}$ gravity for AdS$_4$ gravity, we derive chiral locally AdS$_4$ solutions in the Newman-Unti gauge consistent with a variational principle whose asymptotic symmetry algebra we show, to be an infinite-dimensional chiral extension of $\mathfrak{so}(2,3)$. This symmetry algebra coincides with the chiral $\mathfrak{bms}_4$ algebra in the flat space limit with the corresponding solutions mapping to the space of gravitational vacua in $\mathbb R^{1,3}$ gravity.  We posit this symmetry algebra as the chiral version of recently discovered $\Lambda$-$\mathfrak{bms}_4$ algebra. We propose line integral charges from the bulk AdS$_4$ gravity associated with this asymptotic symmetry algebra and show that they obey the semi-classical limit of a $\mathcal{W}$-algebra. We derive this $\mathcal W$-algebra for finite central charge $c$ and level $\kappa$ using associativity constraints of $2d$ CFT and find it to be isomorphic to  one of  quasi-superconformal algebra that existed in the literature .}

\maketitle

\section{Introduction}
The role of symmetries in physics cannot be over-emphasised. Particularly in the context of two dimensional conformal field theories the symmetry algebras being typically infinite dimensional lead to complete determination of the correlation functions of primary operators. However, it is not always easy to find out the exact symmetry algebra of a given theory.  It is, therefore, important to know what the possible symmetry algebras of a CFT are extending beyond the usual Virasoro algebra. Exploration of this question was initiated by Zamolodchikov long ago \cite{Zamolodchikov:1985wn} in the context of chiral conformal field theories which led to the discovery of new (typically higher spin) extensions of conformal algebra, called the ${\cal W}$-algebras (see \cite{Bouwknegt:1992wg} for a comprehensive review of such algebras). They arise in various contexts in physics including the constrained WZW models. Some of these (at least their classical limits) have been re-discovered using AdS$_3$/CFT$_2$ in more recent times (see, for instance, the initial works {\cite{Henneaux:2010xg}, \cite{Campoleoni:2010zq}). It is also worth noting that all the holographic realisations of  $\mathcal{W}$-algebra have been so far in AdS$_3$/CFT$_2$ and $3d$ Chern-Simons theory set-up and their relevance in $4d$ AdS or flat gravity is still unclear.

In a completely different context, a novel chiral extension of the Poincare algebra $\mathfrak{iso}(1,3)$ has been uncovered recently from the studies of soft-theorems of MHV graviton amplitudes in $4d$ \cite{Banerjee:2020zlg, Banerjee:2021dlm}. This algebra consisted of six chiral operators,  
\bea
\label{currents}
\{ T(z), J_a(z), G_s(z) \} ~~{\rm for}~~ a\in \{0, \pm 1\} ~~{\rm and}~~ s \in \{\pm 1/2 \}.
\eea
The operator $T(z)$ is the usual $2d$ CFT chiral stress tensor with conformal dimension $h=2$ and central charge $c$. The $J_a(z)$ are level-$\kappa$ $\mathfrak{sl}(2,{\mathbb R})$ currents that are conformal primaries with $h=1$ and their Ward identities are equivalent to subleading soft graviton theorem. Finally, the $G_s(z)$ are conformal primaries with $h=3/2$ as well as a doublet of $\mathfrak{sl}(2,{\mathbb R})$ current algebra primaries, and their Ward identities are equivalent to leading soft graviton theorem. The OPE of two $G_s(z)$'s is non-singular, which implies that the supertranslation generators commute. The commutator algebra between the generators is,
 \begin{align}
	&\left[L_m , L_n \right] = (m-n) L_{m+n}\,,~~\left[J_{a,m},J_{b,n}\right] =(a-b)\,J_{a+b,m+n}\,,~~\left[L_m,J_{a,n}\right]=
	-n\,J_{a,m+n},\nonumber\\
	&\left[L_n,P_{s,r}\right]=\frac{1}{2}(n-2r)P_{s,n+r}\,,~~\left[J_{a,m},P_{s,r}\right]=\frac{1}{2}(a-2s)P_{a+s,m+r}\,,~~\left[P_{s,r} ,P_{s',r'} \right] = 0 \label{comm}
\end{align}
We will refer to the algebra (\ref{comm}) as chiral $\mathfrak{bms}_4$ in this paper. In \cite{Gupta:2021cwo}, we proposed chiral boundary conditions for $4d$ asymptotically flat solutions in $\mathbb{R}^{1,3}$ gravity consistent with a variational principle. We found that six holomorphic functions in eq. (\ref{currents}) constitute the full solution, and the asymptotic symmetry algebra of such solutions obey chiral $\mathfrak{bms}_4$ algebra. These locally flat solutions parametrise the space of gravitational vacua and the holomorphic functions in \eqref{currents} can be thought of as Goldstone modes associated with the spontaneous symmetry breaking of chiral $\mathfrak{bms}_4$ symmetry algebra in the vacuum. The chiral boundary conditions used in \cite{Gupta:2021cwo} to obtain $\mathfrak{bms}_4$ algebra form the subset of boundary conditions that give rise to generalised $\mathfrak{bms}_4$ $(\mathfrak{gbms_4})$ as the symmetry algebra of $4d$ asymptotically flat spacetimes. The $\mathfrak{gbms_4}$ algebra is a direct sum of abelian supertranslations  parameterised by an arbitrary function defined on the celestial sphere at null infinity and algebra of vector fields on the celestial sphere $\mathfrak{diff}(\mathbb{S}^2)$. It is a non-chiral extension of Poincare algebra $\mathfrak{iso}(1,3)$ along with extended $\mathfrak{bms}_4$. 
\\
\\
 On the other hand, until recently it was considered that the symmetry algebra of asymptotically AdS$_{d+1}$ spacetimes for $d \geq 3$ is finite-dimensional $\mathfrak{so}{(2,d)}$  for a class of boundary conditions.\footnote{Special cases of infinite dimensional enhancement of symmetry algebra in the case of AdS$_4$ were considered in \cite{Mishra:2017zan, Lowe:2020qan}. In \cite{Taylor:2023ajd}, authors conjectured the existence of deformed $w_{1+\infty}$ for AdS$_4$ spacetime which was then shown to arise from the twistor perspective in \cite{Bittleston:2024rqe}. However the boundary conditions that give rise to  algebra of  \cite{Taylor:2023ajd, Bittleston:2024rqe} in AdS$_4$ gravity in the usual metric formulation is still not known. } Comp\`ere and collaborators in \cite{Compere:2019bua,Compere:2020lrt}, proposed new boundary conditions for generic asymptotically AdS$_4$ spacetimes in Bondi gauge, where they uncover the symmetry algebra of residual gauge transformations to be an infinite dimensional algebra, the $\Lambda$-$\mathfrak{bms}_4$. To be more specific, they fixed only part of the boundary metric using the boundary degrees of freedom. The asymptotic symmetry algebra for such solutions was found to be a Lie algebroid, as the structure constants are background field dependent. The $\Lambda$-$\mathfrak{bms}_4$ and its corresponding phase space in the flat space limit $(\Lambda \rightarrow 0)$ coincide with the generalised $\mathfrak{bms}_4$ algebra and the phase space associated with it. To impose a well-defined variational principle, they further set part of the holographic stress tensor to zero giving rise to reduced symmetry algebra which is a direct sum of $\mathbb{R}\oplus \mathcal{A}$, where $\mathbb{R}$ denotes the abelian time translations and $\mathcal{A}$ is the  algebra of 2-dimensional area-preserving diffeomorphisms.\footnote{This symmetry algebra was also uncovered from Carrollian CFT in \cite{Gupta:2020dtl}}
 
 In this paper, we answer the following two very natural questions that arise due to the existence of chiral $\mathfrak{bms}_4$ in celestial CFT and its realisation from flat space $\mathbb{R}^{1,3}$ gravity,
 \begin{itemize}
 	\item Is there a chiral extension of $\mathfrak{so}(2,3)$ algebra that in some appropriate flat space limit reduces to the chiral $\mathfrak{bms}_4$?
 	\item Can one obtain such a chiral $\Lambda$-$\mathfrak{bms}_4$ algebra from the AdS$_4$ gravity?
 \end{itemize}
To obtain such a chiral extension of  $\mathfrak{so}(2,3)$ from AdS$_4$ gravity, inspired by the chiral boundary conditions in \cite{Gupta:2021cwo}, we propose chiral boundary conditions for AdS$_4$ gravity in Newman-Unti gauge and obtain the symmetry algebra of resultant solutions.

 The boundary conditions that we employ in this paper form a subset of Neumann boundary conditions \cite{Compere:2008us}. We consider locally AdS$_4$ solutions for which the components associated with the holographic stress tensor go to zero and the boundary metric is conformally flat \cite{Skenderis:1999nb}.\footnote{The necessary and sufficient condition for $d=3$ boundary metric to be conformally flat is the vanishing Cotton tensor.} Because the boundary metric is conformally flat, these solutions are referred to as asymptotically AdS$_4$ solutions in the literature \cite{Marolf:2012vvz, Poole:2018koa}. We obtain an infinite-dimensional algebra of asymptotic symmetries associated with these solutions after imposing chiral boundary conditions consistent with a well-defined variational principle. We denote this algebra by chiral $\Lambda$-$\mathfrak{bms}_4$. The chiral $\Lambda$-$\mathfrak{bms}_4$ is a Lie algebroid and in the flat space limit it reduces to the chiral $\mathfrak{bms}_4$ algebra \eqref{comm}. Furthermore, the resultant chiral locally AdS$_4$ solutions that we obtained are in a specific form such that in the flat space limit $(l \rightarrow \infty$, where $l$ is the AdS length), one obtains chiral locally flat solutions derived in \cite{Gupta:2021cwo}. We further argue that the charges associated with this chiral $\Lambda$-$\mathfrak{bms}_4$ symmetry algebra have to be line integral and show that they obey a $\mathcal W$-algebra isomorphic to one of the quasi-superconformal algebra of \cite{Romans:1990ta}.
 The following are the main results of this paper :
\begin{itemize}
	\item  We obtain locally AdS$_4$ solutions parameterised by six holomorphic functions $\{J_{a}(z), T(z),G_{s}(z)\}~~a \in \{0,\pm 1\}, s \in \{\pm \frac{1}{2}\}$ with the algebra of the residual diffeomorphisms being a Lie algebroid. These diffeomorphisms induce non-trivial transformations on the holomorphic functions which are similar to currents in a $2d$ chiral CFT.
	\item  We postulate a line integral charge from AdS$_4$ gravity and obtain the operator product expansions (OPEs) between the currents (of the previous paragraph) which produce the variations of these currents. These OPEs (and the mode commutation relations obtained from them) satisfy a  $\mathcal{W}$-algebra in a semi-classical limit.
	\item We derive the quantum version of this aforementioned $\mathcal{W}$-algebra using tools of $2d$ chiral CFT by imposing Jacobi identities on the commutators of the modes of the chiral (quasi-) primaries, a method developed by Nahm et al. (\cite{Blumenhagen:1990jv} and references therein) as an equivalent implementation of constraints from OPE associativity \cite{Belavin:1984vu}.\footnote{See \cite{DiFrancesco:1997nk}, \cite{Blumenhagen:2009zz} and also the comprehensive review \cite{Thielemans:1994er} of everything to do with OPEs.} We denote this quantum algebra by $ \mathcal{W}(2;(3/2)^2,1^3)$ and show that in the semi-classical limit with proper identification of parameters, this algebra matches with the chiral $\mathcal{W}$-algebra obtained from AdS$_4$ gravity.
	\item The full commutator algebra of $ \mathcal{W}(2;(3/2)^2,1^3)$  is given as follows,
	\begin{align}
		\label{RESULT}
		[L_m, L_n] &= (m-n) \, L_{m+n} + \frac{c}{12} m (m^2-1) \delta_{m+n,0} \, ,\cr
		[L_m, J_{a,n}] &= -n \, J_{a, m+n}, ~~~ [L_m, G_{s,r}] = \frac{1}{2}(n-2r) \, G_{s,n+r} \,,\cr
		[J_{a,m}, J_{b,n}] &= -\frac{1}{2} \kappa \, m \, \eta_{ab} \, \delta_{m+n, 0} + {f_{ab}}^c \, J_{c,m+n}, ~~ [J_{a,n}, G_{s,r}] = G_{s',n+r} {(\lambda_a)^{s'}}_s \, ,\cr
		\!\!\!\!\! [G_{s,r}, \, G_{s',r'}] \!&= \!\epsilon_{ss'}  \Big[\alpha \Big(r^2 - \frac{1}{4} \Big)  \delta_{r+r',0} \! + \! \beta \,  L_{r+r'}  \! + \! \gamma \, (J^2)_{r+r'} \Big] \!\!+ \!\delta (r-r')  J_{a, r+r'}  {(\lambda^a)}_{ss'}  \cr & \cr
		{\rm with} ~~c = &  - \frac{6 \kappa \, (1+ 2 \kappa)}{5+2\kappa},~ \alpha = - \frac{1}{4} \gamma \kappa (3+2\kappa), ~ \beta =  \frac{1}{4}\gamma(5+2\kappa), ~ \delta = -\frac{1}{2} \gamma (3+2\kappa) \cr &
	\end{align}
	where $\eta_{ab} = (3a^2-1) \delta_{a+b,0}$, ${f_{ab}}^c = (a-b) \delta^c_{a+b}$, $(\lambda^a)_{ss'} = \frac{1}{2} \delta^a_{s+s'}$, $(\lambda_a)_{ss'} = \eta_{ab} (\lambda^b)_{ss'}$, $\kappa \ne -5/2$, and $\gamma$ can be fixed to be any non-zero function of $\kappa$ by rescaling the $G_{s,r}$ appropriately. Finally $(J^2)_n$ are the modes of the normal ordered quasi-primary $\eta^{ab} (J_aJ_b)(z)$.  This algebra  is isomorphic to the $N=1$ case of an infinite series of algebra defined for $2N$ bosonic spin-$\frac{3}{2}$ currents and spin-$1$ Kac-Moody current for $\mathfrak{sp}(2N)$ derived by Romans in a different context \cite{Romans:1990ta}.
\end{itemize}   
The rest of this paper is organised as follows: In Section \ref{lads4} we present the calculation of locally AdS$_4$ solution and show the emergence of chiral $\mathcal{W}$-algebra as the asymptotic symmetry algebra after imposing chiral boundary conditions. We then show that the algebra of charges that generate this symmetry algebra is a $\mathcal W$-algebra. In Section \ref{sec2} we derive the quantum version of this $\mathcal W$-algebra from standard $2d$ CFT techniques. We implement the OPE associativity of operator product algebra through the Jacobi Identities of modes of symmetry generating currents of this $\mathcal W$-algebra. We conclude in section \ref{sec7}.

\section{Locally AdS$_4$ solutions}\label{lads4} 
To obtain  chiral $\Lambda$-$\mathfrak{bms}_4$ algebra from AdS$_4$ gravity, we will analyse the asymptotic symmetries of locally AdS$_4$ geometries (lAdS$_4$). We work in Newman-Unti (NU) gauge with the coordinates $(u,r,z,\bar{z})$. We prefer this gauge over the most commonly used Fefferman-Graham (FG) gauge because it is straightforward to compare lAdS$_4$ solutions with the locally flat solutions of \cite{Gupta:2021cwo} in NU gauge. This gauge is defined by imposing the following conditions on the metric $g_{\mu \nu}$,
\begin{align}
	g_{rr} =g_{rz} = g_{r\bar z}=0, ~~~ g_{ur} =-1 .
\end{align}
The line element is then given by,
\begin{align}
	ds^2_{lAdS}=-2\,du\,dr+ g_{ij}\,dx^i\,dx^j , \label{line element}
\end{align}
where $r$ is the radial coordinate and $i$ labels the boundary directions. The metric components $g_{ij}$ for $i,j \in \{ u, z, \bar z\}$ are expanded in the following form:
\begin{align}
	g_{ij} (r, u, z, \bar z) = \sum_{n=0}^\infty r^{2-n} \, g_{ij}^{(n)}(u,z, \bar z) .
\end{align}
We seek geometries, which in addition to being the solution to Einstein equation for negative cosmological constant in four dimensions,
\begin{align}
	E_{\mu \nu}= R_{\mu \nu}+ \frac{3}{l^2}\,g_{\mu \nu}=0 \label{eom}
\end{align}
also satisfies locally \text{AdS$_4$} condition given by
\begin{align}
	A_{\mu\nu\sigma\lambda}=R_{\mu\nu\sigma\lambda}+\frac{1}{l^2}\,(g_{\mu\sigma}g_{\nu \lambda}-g_{\mu \lambda}g_{\nu \sigma})=0. \label{Lads cond1}
\end{align}
Equation (\ref{Lads cond1}) implies equation of motion (\ref{eom}). These lAdS$_4$ solutions share the same boundary conditions with solutions to asymptotically locally AdS$_4$ condition given by ,
\begin{align}
	R^{\mu}_{~\nu\sigma\lambda}+\frac{1}{l^2}\,(\delta^{\mu}_{\sigma}\,g_{\nu \lambda}-\delta^{\mu}_{\lambda}\,g_{\nu \sigma}) \rightarrow 0 \label{Ladscond2}
\end{align}
as $r\rightarrow \infty$ following the analysis of \cite{Gupta:2021cwo}. Taking the flat space limit $l \rightarrow \infty$ of \eqref{Lads cond1} and \eqref{Ladscond2}, we obtain the conditions for locally flat  and asymptotically locally flat  geometries that was used in our earlier work \cite{Gupta:2021cwo}. Using the coordinate transformations it is straightforward to map the solution in NU gauge to FG gauge as shown in \cite{Compere:2019bua, Compere:2020lrt}. 
 Imposing the condition (\ref{Lads cond1}) one finds that the solution terminates at order $\mathcal{O}(1)$ in power series expansion in $r$ such that for locally AdS$_4$ solutions in $4d$ we have,
\begin{align}
	g_{ij} (r, u, z, \bar z) = \sum_{n=0}^2 r^{2-n} \, g_{ij}^{(n)}(u,z, \bar z). 
\end{align}

We begin by solving Einstein equation  $E_{\mu \nu}=0$ for the solution ansatz (\ref{line element}) in NU gauge at each order in $r$ near the boundary $(r \rightarrow \infty)$. From $\mathcal{O}(r^2)$ term of the component $E_{uu}$ one obtains the condition,
\begin{align}
	Det\, (g^{(0)}_{ij})=-\frac{1}{l^2}\,Det\,(g^{(0)}_{ab})\,~~ a \in \{z,\bar{z}\},\label{det cond}
\end{align} 
where $a$ labels the index on codimension-two hypersurface. This condition is solved by taking $g^{(0)}_{uu}$ to be of the form,
\begin{align}
	g^{(0)}_{uu}= -\frac{1}{l^2}+ v_a\,v^a ,
\end{align}
where we define $v_a \coloneqq g^{(0)}_{u a} $ in the paper and the indices $\{a,b\}$ are lowered and raised by boundary metric $g^{(0)}_{ab}$ and $g_{(0)}^{ab}$ respectively.
The algebraic condition at $\mathcal{O}\left(\frac{1}{r}\right)$ in $E_{ra}$ gives the expression for $g^{(1)}_{ua}$ as follows,
\begin{align}
	g^{(1)}_{u a}= g^{(1)}_{ab}\,v^b \label{gu1}.
\end{align}
Similarly the order $\mathcal{O}\left(\frac{1}{r}\right)$ term of $E_{ur}$ and $E_{r a}$  can be solved respectively to obtain $g^{(1)}_{uu}$ and $g^{(2)}_{ua}$,
\begin{align}
	&g^{(1)}_{uu} = -\frac{\theta}{2}+ D_a v^a- 2\,g^{(1)}_{ab}\,v^a\,v^b+ 3\,v^a g^{(1)}_{u a}-\frac{1}{2 l^2}\, g_{(0)}^{ab}\,g^{(1)}_{ab}, \\
	&	g^{(2)}_{ua}=\frac{1}{2}g^{(2)}_{ab}v^b+ \frac{1}{2}\,g_{(0)}^{cb}\,D_{c}g^{(1)}_{ab}-\frac{1}{2}\,g_{(0)}^{cb}\,D_{a}\,g^{(1)}_{cb},
\end{align}
where $\theta= g_{(0)}^{ab}\,\partial_u g^{(0)}_{ab}$ and $D_a v^a$ is the covariant derivative defined with respect to the $2d$ boundary metric $g^{(0)}_{ab}$. 
At $\mathcal{O}\left(r\right)$ in $E_{ab}$ we obtain the following condition that will be used to solve for $g^{(1)}_{ab}$,
\begin{align}
	\frac{1}{l^2\,}g^{(1)}_{ab}-\bigg[D_a\, v_b+ D_b\,v_a - D_c v^c\,g^{(0)}_{ab}+ \frac{1}{2}\,\theta g^{(0)}_{ab}- \partial_u g^{(0)}_{ab}\bigg]- \frac{1}{2\,l^2}\, g_{(0)}^{cd}\,g^{(1)}_{cd}\,g^{(0)}_{ab}=0. \label{g1 cond}
\end{align} 
 The $\mathcal{O}(1)$ term in $E_{ab}$  gives us the following condition,
\begin{align}
	&g^{(0)}_{ab}\,\bigg[g^{(2)}_{uu}+ \frac{R^{(0)}}{2}-\frac{1}{4 \,l^2}\,g^{(1)}_{cd}\,g_{(1)}^{cd}+\frac{1}{2} g_{(0)}^{cd}\,\partial_u g^{(1)}_{cd}- \frac{1}{4} \,g_{(0)}^{ab}\,g^{(1)}_{ab}\,\theta- \frac{1}{2}\,g_{(0)}^{cd}\,v^e\,D_e\,g^{(1)}_{cd} \nonumber \\ &+ v^c\,v^d\,g^{(2)}_{cd}+ \frac{1}{l^2}\, g_{(0)}^{ab}\,g^{(2)}_{ab}- 2\,g^{(2)}_{u c}\,v^c + \frac{1}{2}\, g_{(0)}^{ab}\,g^{(1)}_{ab} D_c v^c- g_{(1)}^{cd} D_c v_d\bigg] \nonumber \\&+ \frac{1}{4}\, Tr \,g_{(0)}^{ab}\,g^{(1)}_{ab}\,\partial_u g^{(0)}_{ab}+ \frac{1}{l^2} g^{(2)}_{ab}- \frac{1}{2}\, g_{(0)}^{ab}\,g^{(1)}_{ab}\,D_{(a}\,v_{b)}=0
\end{align}
where $R^{(0)}$ is the Ricci scalar of the boundary metric $g^{(0)}_{ab}$.
Taking the trace of this equation allows us to solve for $g^{(2)}_{uu}$,
\begin{align}
	&g^{(2)}_{uu}=-\frac{R^{(0)}}{2}+\frac{1}{4\,l^2}\, g^{(1)}_{ab}g_{(1)}^{ab}-\frac{1}{2}\,g_{(0)}^{ab}\,\partial_u g^{(1)}_{ab}+\frac{1}{8}\,g_{(0)}^{ab}\,g^{(1)}_{ab}\,\theta+\frac{1}{2}\,g_{(0)}^{ab}\,v^e\,D_e\,g^{(1)}_{ab}-v^a\,v^b\,g^{(2)}_{ab} \nonumber \\&-\frac{3}{2\,l^2}\,g_{(0)}^{ab}\,g^{(2)}_{ab} + 2\,g^{(2)}_{u a}\,v^a-\frac{1}{4}\,g_{(0)}^{cd}\,g^{(1)}_{cd}\,D_a\,v^a+ g_{(1)}^{ab}\,D_a v_b.
\end{align}

In order to obtain $g^{(2)}_{ab}$, we impose locally AdS condition of (\ref{Lads cond1}) which is easier to implement in the form
\begin{equation}
	\label{modlads}
	\det({g^{(4d)}})\,A_{\mu\nu\rho\sigma}=0,
\end{equation}
as the series expansion in $r$ is a positive power with highest order 8. $\det({g^{(4d)}})$ is the determinant of full $4d$ metric in NU gauge. The $\mathcal{O}(r^2)$ term of component $\det({g^{(4d)}})\,A_{urur}$ gives $g^{(2)}_{ab}$ as follows,
\begin{align}
	&g^{(2)}_{ab}= \frac{1}{4}g^{(1)}_{ac}\,g^{cd}_{(0)}\,g^{(1)}_{db}. 
\end{align}
The pullback of the metric in the Newman-Unti gauge at the boundary $\mathcal{I}$ at $r\rightarrow \infty$ is given by,
\begin{align}
	ds^2_{bdry}=\left(-\frac{1}{l^2}+v_av^a\right)du^2+2\,v_a\,du\,dx^a+g^{(0)}_{ab}dx^a dx^b. \label{bdrymetric}
\end{align}
Some comments about our solutions are in order, 
\begin{enumerate}
	\item In the flat space limit the condition (\ref{det cond}) implies that $g_{ij}$ is degenerate and another way to solve it is to set $g^{(0)}_{uu}= g^{(0)}_{uz}= g^{(0)}_{u \bar{z}}=0$ as we did in \cite{Gupta:2021cwo}. 
	\item The condition in equation (\ref{g1 cond}) suggest that $g^{(1)}_{ab}$ is not a free data on the boundary and is given in terms of $v_a$ and $g^{(0)}_{ab}$. This is in contrast to locally flat solutions where $g^{(1)}_{ab}$ is a free data at the boundary null infinity. This can be seen from taking the flat limit of (\ref{g1 cond}) where the term proportional to $g^{(1)}_{ab}$ drops out and after putting $v_a=0$ one is left with the condition constraining the time dependence of the boundary metric $g^{(0)}_{ab}$. For locally flat solutions this equation implied that the boundary metric is conformally time dependent i.e $(g^{(0)}_{ab}= \Omega(u,z,\bar{z})\,q_{ab}(z,\bar{z}))$ where we choose $\Omega$ to be $u$-independent such that $\theta=0$. For lAdS$_4$ solutions we do not have any such restriction on $g^{(0)}_{ab}$. 
	\item We further set $g_{(0)}^{ab}g^{(1)}_{ab}=0$ as part of our gauge condition. This is standard practice for solutions in NU and Bondi gauge and is equivalent to the condition of fixing the origin of affine parameter of null geodesic \cite{Barnich:2011ty}. The final  metric components are then given by,
	\begin{equation}
		\begin{split}
			&g^{(1)}_{uu} = -\frac{\theta}{2}+ D_a v^a+ g^{(1)}_{ab}\,v^a\,v^b \\
			&g^{(2)}_{ua}=\frac{1}{2}g^{(2)}_{ab}v^b+ \frac{1}{2}\,g_{(0)}^{cb}\,D_{c}g^{(1)}_{ab}\\
			&g^{(1)}_{ab}=l^2\,\bigg[D_a\, v_b+ D_b\,v_a - D_c v^c\,g^{(0)}_{ab}+ \frac{1}{2}\,\theta g^{(0)}_{ab}- \partial_u g^{(0)}_{ab}\bigg]\\
			&g^{(2)}_{uu}=-\frac{R^{(0)}}{2}-\frac{1}{8\,l^2}\, g^{(1)}_{ab}g_{(1)}^{ab}-\frac{1}{2}\,g_{(0)}^{ab}\,\partial_u g^{(1)}_{ab}+ g_{(0)}^{bc}\,v^a\,D_c\,g^{(1)}_{ab} + g_{(1)}^{ab}\,D_a v_b \label{fullsol}
		\end{split}
	\end{equation}
	along with equations (\ref{det cond}) and (\ref{gu1}).
	\item One can check that for $v_a=0$ in the limit $l \rightarrow \infty$ the metric components coincide with those  of locally flat solutions in \cite{Gupta:2021cwo} except that  $g^{(1)}_{ab}$ is free data.
\end{enumerate}
For $\text{lAdS}_4$ solutions the boundary metric $g^{(0)}_{ab}$ and $v_a$ are the free data on boundary,\footnote{Note that $\{g^{(0)}_{ab}, v^a\}$ constitute only five independent metric components of $3d$ boundary metric. The sixth component is fixed due to the condition \eqref{det cond}.} in terms of which all other metric components are given. For our solutions to completely solve equations  (\ref{eom}) and (\ref{Lads cond1}), the Cotton tensor of the boundary metric \eqref{bdrymetric} necessarily vanishes following the analysis of \cite{Skenderis:1999nb}. This imposes constraints on boundary data $g^{(0)}_{ab}$ and $v_a$. We do not provide these constraints nor their general solution in this paper as it is not needed for our analysis. We will solve the constraints only after imposing further boundary conditions on our free data $\{g^{(0)}_{ab}, v_a\}$, which we obtain after imposing a well defined variational principle.
\subsection{Variational principle}
In this section we will find boundary conditions for the fields in configuration space motivated by a variational principle which will then allow us to  completely solve the constraint conditions on the free data. The action for Einstein equation in $4d$ for non-zero cosmological constant is given by,
\begin{align}
	S= \frac{1}{16 \pi G}\int d^4 x \,\sqrt{-g}\,\big( R+ \frac{6}{l^2}\big)
\end{align}
The variation of this action gives equation of motion and a boundary term which depends on the normal derivative of the boundary metric. To solve the variational principle one usually adds Gibbons Hawking term and various counter terms such that in the end total variation of the action is only proportional to the boundary component and not its derivative. This is the usual procedure when one works in FG gauge. As we are working in NU gauge, we use the boundary terms that we found in \cite{Gupta:2021cwo} for the case of asymptotically locally flat spacetimes in NU gauge.\footnote{Normally the boundary action for AdS$_{d+1}$ gravity in Fefferman-Graham gauge \cite{Balasubramanian:1999re, Brown:1992br} or the gauge used to describe `boosted black brane' \cite{Bhattacharyya:2007vjd} is invariant under $d$-dimensional boundary preserving diffeomorphisms unlike the boundary terms used here in context of NU gauge. We expect the variational problem that gives rise to chiral symmetry algebra in NU gauge to hold for these gauges as well. We leave a comparison of these boundary terms with that of standard Gibbon-Hawking term and other counterterms for future work.} The derivation of finding boundary terms for locally AdS$_4$ follows identically to \cite{Gupta:2021cwo} and therefore we use only the final answer here. The boundary action that we add to Einstein-Hilbert action is,
\begin{align}
&&S_{bdry}=(16 \pi G)^{-1} \sqrt{\sigma} \left[ 2 \,  \, V^a \partial_r  V_a -  \partial_r \omega  - \left (\omega \, \sigma^{ab} +  V^a V^b - V^c V_c \, \sigma^{ab} \right) \, \partial_r \sigma_{ab}   -  \sigma^{ab} \, \partial_u  \sigma_{ab} \right] \cr
	&& ~~~~~~~~~~ = \sqrt{\sigma} \, \partial_r  (V^a V_a - \omega) +2 \, \left[ (V^a V_a - \omega ) \,  \partial_r - \partial_u \right] \, \sqrt{\sigma}
\end{align}
and the total divergence term in $\delta S_{EH+bdry}$ is $(16 \pi G)^{-1}$ times:
\begin{align}
	\partial_a \left[\sqrt{\sigma} \, \left( \delta V^a - V^a \, \sigma_{bc} \, \delta \sigma^{bc} \right) \right]
\end{align}
that we ignore (this amounts to assuming that the geometry of $\Sigma_2$ with coordinates $(z, \bar z)$ is either compact or, when it is not, the integrand falls-off fast enough near its asymptotes). Here $V_a= g_{ua}\,,\omega= g_{uu}\,,\sigma_{ab}= g_{ab}$. Substituting the series expansion of these components near the boundary $\mathcal I$ one obtains the following expansion of the variation of total Lagrangian density after adding boundary action,  
{\small
	\begin{align}
		& \delta {\cal L}_{EH+bdry} = -2 \, r^3 \, \sqrt{g^{(0)}} \, \delta g_{uu}^{(0)} - r^2 \sqrt{g^{(0)}} \, \left[ \frac{1}{2} \left(g_{(0)}^{cd} g^{(1)}_{cd} \right) \, \delta g_{uu}^{(0)} + 2 \, \delta g_{uu}^{(1)} \right] \cr
		& - r \, \sqrt{g^{(0)}} \, \Big[2 \, \delta g_{uu}^{(2)} + g_{uu}^{(2)} \, g_{(0)}^{ab} \, \delta g^{(0)}_{ab} + \frac{1}{2} \left(g_{(0)}^{cd} g^{(1)}_{cd} \right) \, \delta g_{uu}^{(1)} - 2 \, g^{(2)}_{ua} g_{(0)}^{ab} \delta g^{(0)}_{ub}  \cr
		& ~~~~~~~~~~~~~~~~ - \frac{1}{2} \left( g_{(0)}^{ab}\partial_u g^{(1)}_{bc} g_{(0)}^{cd} \, \delta g^{(0)}_{da} - \left(g_{(0)}^{cd} \partial_u g^{(1)}_{cd} \right) \left(g_{(0)}^{ab} \, \delta g^{(0)}_{ab} \right) \right]  - \sqrt{g^{(0)}} \,\,  \delta {\cal L}_0 + {\cal O}(1/r)  \cr & \label{var1}
	\end{align}
}
where:
\begin{align}
	& \delta {\cal L}_0 =  \frac{1}{2} \left(g_{(0)}^{cd}  g^{(1)}_{cd} \right) \, \delta g_{uu}^{(2)} + \frac{1}{2} g_{uu}^{(2)} \left(g_{(1)}^{ab} \delta g^{(0)}_{ab} \right) 
	\cr
	& - \frac{1}{2} \, \Big[\frac{1}{2}  \rm Tr\,(g_{(0)}^{-1}\, g^{(1)}) \,\rm Tr\,(g_{(0)}^{-1} g^{(2)}) - \rm Tr \left( g^{(1)} g_{(0)}^{-1} g^{(2)} g_{(0)}^{-1} \right) \cr
	& - \frac{1}{8} \rm Tr\,(g_{(0)}^{-1}\, g^{(1)}) \rm Tr \left( g^{(1)} g_{(0)}^{-1} g^{(1)} g_{(0)}^{-1} \right) + \frac{1}{4} \rm Tr \left( g^{(1)} g_{(0)}^{-1} \right)^3 \Big] \, \delta g_{uu}^{(0)} -  \left(D_{(a} g^{(2)}_{b)u}- D^c g_{uc}^{(2)} \, g^{(0)}_{ab} \right) \, \delta g_{(0)}^{ab}\cr
	& - \frac{1}{4} \rm Tr(g_{(0)}^{-1}\, g^{(1)}) \, \left(\partial_u g^{(1)}_{ab} - \partial_u (\rm Tr g^{(1)}) g^{(0)}_{ab} \right) \, \delta g_{(0)}^{ab} +\frac{1}{2} \left(  \partial_u g^{(2)}_{ab} \delta g_{(0)}^{ab} - \rm Tr (g^{-1}_{(0)}\,\partial_u g^{(2)}) \, \left( g^{(0)}_{ab} \, \delta g_{(0)}^{ab} \right) \right) \cr
	& + g^{(2)}_{ua}  g_{(1)}^{ab}  \delta g^{(0)}_{bu}  - 2 g^{(2)}_{au} g_{(0)}^{ab} \, \delta g^{(1)}_{bu} + g^{(2)}_{uu} \, g_{(0)}^{ab} \delta g^{(1)}_{ab} \cr
	& - \frac{1}{2} \left( \rm Tr \left( g_{(0)}^{-1} \partial_u g^{(1)} g_{(0)}^{-1} \delta g^{(1)}\right) - \rm Tr \left( g_{(0)}^{-1} \partial_u g^{(1)} \right) \rm Tr \left(g_{(0)}^{-1} \delta g^{(1)}\right)  \right) \label{var4}
\end{align}
with ${\rm Tr} \left(g_{(0)}^{-1}\,g^{(2)}\right) = g_{(0)}^{ab} g^{(2)}_{ab}$ and so on. The variation is calculated at  $r=r_{0}$ hypersurface and in the end we take the limit $r_{0} \rightarrow \infty$. The terms at $\mathcal{O}\left(\frac{1}{r^n_{(0)}}\right)$ for $n\geq 1$ vanish in this limit and therefore one does not need to consider these terms for our analysis.

Now we impose boundary conditions on the boundary data such that the total variation in (\ref{var1}) is zero on the solution space.
\begin{enumerate}
	\item For terms at $\mathcal{O}(r^3)$ and $\mathcal{O}(r^2)$ to vanish one has to impose $\delta g^{(0)}_{uu}= \delta g^{(1)}_{uu}=0$. To satisfy these conditions, we set $v_a = \delta v_a =0$ and $\delta \theta=0$. The latter condition is solved by keeping determinant of the $2d$ spatial boundary metric fixed such that $\delta \sqrt{det\,g^{(0)}_{ab}}=0$. In order to impose such a condition we parameterise our boundary metric $g^{(0)}_{ab}$ in chiral Polyakov gauge where we set $g^{(0)}_{\bar{z}\bar{z}}=0$ and $g^{(0)}_{z \bar{z}}=\Omega (z,\bar{z})$ and $\delta g^{(0)}_{\bar{z}\bar{z}}=\delta g^{(0)}_{z\bar{z}}=0$. $\Omega (z,\bar{z})$ is the conformal factor of $2d$ metric which is $\frac{1}{2}$ for $\mathbb{R}^2$ and $\frac{4}{(1+ z\bar{z})^2}$ for $\mathbb{S}^2$. In rest of the analysis we work with $\Omega(z,\bar{z})= \frac{1}{2}$. It is straightforward to extend the solutions for the case of $\mathbb{S}^2$. The line element of resultant $2d$ boundary metric is given as follows,
	\begin{align}
		ds^2_{2d}=g^{(0)}_{zz}(u,z,\bar{z}) \,dz^2+ dz\,d\bar{z} \label{co-dimension-two}
	\end{align} 
These chiral boundary conditions on the boundary metric is similar to what was imposed in \cite{Gupta:2021cwo} except now the boundary metric has $u$ dependence as well.\footnote{These chiral boundary conditions are similar to boundary conditions of AdS$_3$ gravity used in \cite{Avery:2013dja}. } 
	\item At order $\mathcal{O}(r)$ we impose one more condition which is,
	\begin{align}
		\delta \int d^2 z\,\bigg[\sqrt{g^{(0)}}\,R^{(0)}\bigg]. \label{varcond1}
	\end{align}
	This is equivalent to holding the Euler Character of the boundary metric $g^{(0)}_{ab}$ fixed. This condition was also used in \cite{Compere:2018ylh}. Using \eqref{varcond1} and the conditions obtained in previous point, one can check that $\mathcal O(r)$ terms in \eqref{var1} vanish.
	\item Using the boundary conditions mentioned in the previous two points, along with the gauge condition $g_{(0)}^{ab}\,g^{(1)}_{ab}=0$ we choose to impose (as mentioned before), one can see that the $\mathcal O(1)$ terms denoted by $\delta \mathcal L_0$ in \eqref{var1}  also vanish without imposing any additional conditions. 
\end{enumerate} 
We impose these boundary conditions on the configuration space to ensure that $\delta \mathcal{L}_{EH+bdy}$ vanishes on the solution space that share the same boundary conditions.  In literature the boundary conditions  $v_a = \delta v_a = \delta \sqrt{det\,g^{(0)}_{ab}}=0$ were also imposed in \cite{Compere:2019bua} where the asymptotic symmetry algebra of asymptotically locally AdS$_4$ was found to be $\Lambda$- $\mathfrak{bms}_4 $. These conditions were realised using the gauge freedom of the boundary diffeomorphisms and Weyl scaling of boundary metric data. Instead of imposing chiral boundary conditions their treatment kept the boundary metric $g^{(0)}_{ab}$ general with determinant fixed.

Once we impose these boundary conditions on our configuration space we see that  $g^{(0)}_{zz}$ is the only free data on the boundary. One can now easily solve (\ref{eom}) and (\ref{Lads cond1}) completely by solving the following constraint equations that are obtained from the $\mathcal{O}(r^3)$ terms of $\det({g^{(4d)}})\,A_{uru\bar{z}}$, $\det({g^{(4d)}})\,A_{uruz}$ and  $\mathcal{O}(r^5)$ terms of $\det({g^{(4d)}})\,A_{uzuz}$ respectively,
\begin{align}
	&\partial_u\,\partial_{\bar{z}}^2\,g^{(0)}_{zz}=0, ~~ \partial_z\,\partial_{\bar{z}}^2\,g^{(0)}_{zz}= l^2\,\partial_u^2\,\partial_{\bar{z}} g^{(0)}_{zz} \label{lAdscon1} \\
	& \partial_u\,\bigg[\frac{1}{2}\,\big(\partial_{\bar{z}}g^{(0)}_{zz}\big)^2-g^{(0)}_{zz}\,\partial_{\bar{z}}^2\,g^{(0)}_{zz}+ \partial_z\,\partial_{\bar{z}}\,g^{(0)}_{zz}- \frac{l^2}{2}\partial_u ^2\,g^{(0)}_{zz}\bigg]=0. \label{lAdscon2}
\end{align}
It is important to keep track of the correct powers of $l$ in the solutions because  $g^{(0)}_{ab}$ should have no length dimension and hence any power of $u$ should be accompanied with $\frac{1}{l}$ so as to make it dimensionless. Being careful about correct scaling factor plays an important role while taking the flat space limit. Since $g^{(1)}_{ab}$ has length dimension one and should also survive the flat space limit, we write down the solutions in a way that it scales as $\sqrt{G}$ ($G$ is the Newton constant) in the limit $l \rightarrow \infty$.\footnote{One can also use Planck length $l_{pl}$.} Keeping these things in mind, the solution to the equations (\ref{lAdscon1}) and (\ref{lAdscon2}) is given by,
\begin{align}
	g^{(0)}_{zz}= \Gamma(u,z)+J_{-1}(z)+ \bar{z}^2\,J_{1}(z)+ \bar{z}\,\bigg(J_{0}(z)- \frac{\sqrt{G}\,u}{l^2}G_{1/2}(z)+ \frac{u^2}{l^2}\,\partial J_1(z)\bigg) \label{co-dimension two 2}
\end{align}
\begin{multline}
	\Gamma(u,z)= \frac{u\,\sqrt{G}}{l^2}\,\bigg(\frac{3\,G_{1/2}(z)\,\partial J_1(z)}{4\,J_{1}(z)^2}-\frac{\partial G_{1/2}(z)}{2\,J_1(z)}-\frac{G_{1/2}(z)\,J_0(z)}{2\, J_1(z)}\bigg)  \\ + \frac{u^2}{l^2}\,\left(\,\frac{G\,G_{1/2}(z)^2}{4\,l^2\, J_1(z)}+\frac{J_0(z)\,\partial J_1(z)}{2\,J_1(z)}-\frac{3\,(\partial J_1(z)) ^2}{4\,J_1(z)^2}+ \frac{\partial^2 J_1(z)}{2\,J_1(z)}\,\right)\\-\frac{u^3\,\sqrt{G}\,G_{1/2}\,\partial J_1(z)}{2\,l^4\,J_1(z)}
	+ \frac{u^4\,(\partial J_1(z))^2}{4\,l^4\,J_1(z)}+ \frac{\sin{\left(\frac{2\,u\,\sqrt{J_1(z)}}{l}\right)}}{\sqrt{J_1(z)}}\,K_1(z)+\frac{\sin ^2{\left(\frac{u\,\sqrt{J_1(z)}}{l}\right)}}{J_1(z)}\,K_2(z) \label{gamma}
\end{multline}
where,
\begin{multline}
	K_1(z)=\frac{\sqrt{G}}{8\,l\,J_1(z)^{2}}\,\big[2\,J_1(z)\,\left(\partial\,G_{1/2}(z)-2\,G_{-1/2}(z)\,J_1(z)\right)\\ + G_{1/2}(z)\,\left(2\,J_0(z)\,J_1(z)-3\,\partial J_1(z) \right)\big]
\end{multline}
\begin{multline}
	K_2(z)= -\frac{1}{2}\,\big(\kappa(z)-\frac{1}{2}\,J_{0}(z)^2-\partial J_0 (z)+ 2\,J_1(z)\,J_{-1}(z)\big)\\+\frac{1}{4\,J_1(z)^2}\,\big[3\,(\partial J_1)^2-\,J_0(z)\,\partial(J_1(z)^2)-2\,J_1(z)\,\partial^2\,J_1(z) -\frac{G}{l^2}\,J_1(z)\,G_{1/2}(z)^2\big]	
\end{multline}
Let us discuss some properties of these solutions,
\begin{enumerate}
	\item As mentioned before the only free data is $g^{(0)}_{zz}$. All other metric components are obtained from $g^{(0)}_{zz}$.
	\item These solutions are parameterised by six holomorphic functions,
	\begin{align}
		\{J_a(z)\,,\kappa(z)\,,G_{s}(z)\} ~~{\rm for}~~ a\in \{0, \pm 1\} ~~{\rm and}~~ s \in \{\pm 1/2 \}.\label{funcs}
	\end{align}
	\item As can be seen from the solutions, function $J_{1}(z)$  appears in the denominator at several places. However the solutions have a well defined limit as $J_{1}(z)\rightarrow 0$ along with the other five functions in (\ref{funcs}) such that the solution does not diverge. One gets back to the global AdS$_4$ solution when all the six holomorphic functions approach zero value.
	\item The solution has a well defined flat space limit $(l \rightarrow \infty)$ such that,
	\begin{align}
		ds^2_{LAdS_4}= ds^2_{LF}+ \mathcal{O}\bigg(\frac{1}{l^2}\bigg)+ \cdots
	\end{align}  
	where $\cdots$ indicate higher order terms in $\frac{1}{l^2}$ with, 
	\begin{align}
		\label{finalLFr2}
		ds^2_{LF}=& -\,2\,du\,dr +\,r^2\,dz\,d\bar{z} +r^2\,\big(J_p (z)\,\bar{z}^{1+p}\big)\,dz^2 \cr
		&+r\,u\,\bigg(\,\frac{1}{u}\,\sqrt{G}\,G_s(z)\,\bar{z}^{\frac{2s+1}{2}}+\kappa(z) +\frac{1}{2} J_p (z) J^p (z) -(1+p)\, \bar{z}^p\,\partial_z\,J_p(z)\bigg)\,dz^2 \cr
		& +2 \,\left(\frac{2s+1}{2}\,\sqrt{G}\,G_s\,\bar{z}^{\frac{2s-1}{2}}-\,u\,p\,(1+p)\,\partial_z J_p\,\bar{z}^{p-1}\right)\,du\,dz \cr
		& - 2\,p\,(1+p)\,J_p\,\bar{z}^{p-1}\,du^2, ~~~~ p \in \{-1,0,1\},~ r \in \{-\frac{1}{2},\frac{1}{2}\}
	\end{align}
	The expression of $ds^2_{LF}$ coincides with the locally flat solutions which were found by us in \cite{Gupta:2021cwo}, where the topology of the boundary metric at the boundary of flat space is $\mathbb{R}^2$ with the identification $\{J_p \rightarrow - J_p\,,\kappa(z)\rightarrow -\kappa(z), G_s \rightarrow \frac{C_s}{\sqrt{G}}\,\}$.
\end{enumerate}
Now that we have obtained $\text{lAdS}_4$ solutions with chiral boundary conditions we turn next to calculating the asymptotic symmetry algebra, i.e., a set of residual diffeomorphisms that keep this solution within the same class. In the process we will derive the variations of the six chiral holomorphic functions (\ref{funcs}) that parameterise our solution space under these residual gauge transformations.
\subsection{Asymptotic symmetry algebra }
%

We look for residual gauge transformations that take us from one $\text{lAdS}_4$ solution to another without spoiling the gauge and boundary conditions of our initial solution. In order to preserve the gauge conditions of NU gauge the vector fields  of the form $\xi^\mu \partial_\mu$ should satisfy,
\begin{align}
	\delta g_{ru} = \delta g_{rr}=\,\delta g_{ra}=0 \label{gaugecond1}
\end{align}
The variation in the metric component is the result of the Lie derivative of the metric component along the vector field $\xi^\mu\,\partial_{\mu}$ such that $$\delta g_{\mu \nu}= -\mathcal{L}_{\xi}\,g_{\mu \nu}.$$ The conditions (\ref{gaugecond1}) constraint the vector field components as follows:
\begin{align}
	\label{asv1}
	\partial_r\xi^u = 0, ~~ \partial_u \xi^u + \partial_r \xi^r = g_{ua} \, \partial_r \xi^a, ~~\partial_a \xi^u = g_{ab} \partial_r \xi^b
\end{align}
From first of the equation in (\ref{asv1}) we see that $\xi^u= \xi^u_{(0)}(u,z,\bar{z})$. The other two equations can further be solved in an asymptotic expansion around large-$r$. We expand $\xi^r$ and $\xi^a$ near the boundary as follows,
\begin{align}
	\xi^r = \sum_{n=0}^\infty r^{1-n} \xi^r_{(n)} (u,z,\bar z), ~~ \xi^a = \sum_{n=0}^\infty r^{-n} \xi^a_{(n)} (u,z,\bar z),~~ a \in \{z,\bar{z}\}
\end{align}
At each order in $r$ the equations (\ref{asv1}) are solved recursively to obtain the terms at subleading order in $r$ of $\xi^r$ and $\xi^i$ in terms of leading coefficients $\xi_{(0)}^r$ and $\xi_{(0)}^i $ and $\xi^u_{(0)}$. The first few such terms are given by, 
\begin{equation}
	\begin{split}
		&\xi^r_{(0)}= -\partial_u \xi^u_{(0)}+ v^a\,\partial_a\xi^u_{(0)}\,,~~ \xi^r_{(2)}= g^{(2)}_{ua}\,\xi^a_{(1)}+ 2\,\xi^a_{(2)}\,g^{(1)}_{ua} \\
		& \xi^r_{(n)}= \xi^a_{(n-1)}\,g^{(2)}_{ua}\,,~~ \text{for}~\,n=3,4,\cdots\\
		& \xi^a_{(1)}= -g_{(0)}^{ab}\, \partial_b\xi^u_{(0)}\,,~~
		\xi^a_{(2)}=\frac{1}{2}\,g^{ab}_{(0)}\,g^{(1)}_{bc}\,g^{cd}_{(0)}\,D_d\,\xi^u_{(0)}. \label{vecfield_cond1}
	\end{split}
\end{equation}
Also demanding that the vector fields preserve the traceless condition of $g^{(1)}_{ab}$ allows us to solve for $\xi^r_{(1)}$ which is given by,
\begin{align}
	\label{asv3}
	\xi^r_{(1)} = \frac{1}{2} \square \, \xi^u - \frac{1}{4} \xi^a_{(0)} D_a \left(g_{(0)}^{bc} g^{(1)}_{bc} \right) - \frac{1}{4} \partial_u \left( \xi^u \, g_{(0)}^{bc} g^{(1)}_{bc}  \right)+\frac{1}{4}\,v^c\,\partial_c\,\xi^u_{(0)}\,g^{ab}_{(0)}\,g^{(1)}_{ab}.
\end{align}
Here $\xi^a_{(0)}$ and $\xi^u_{(0)}$ are unconstrained functions of $(u,z,\bar{z})$ and constitute boundary vector fields that are obtained by pull back of $\xi$ on the boundary.
\begin{align}
	\xi_{bdry}= \xi^u_{(0)}\,\partial_u + \xi^a_{(0)}\,\partial_a \label{bdryvec}
\end{align} 
The variation of free data $g^{(0)}_{ab}$ and $v_a$ in terms of these components of boundary vector fields are given by,
\begin{align}
	\delta g^{(0)}_{ab}&= \xi^u_{(0)}\partial_u\,g^{(0)}_{ab}+ \mathcal{L}_{\xi^c_{(0)}}\,g^{(0)}_{ab}+ 2\,v_{(a}\partial_{b)}\,\xi^u_{(0)}-2\,g^{(0)}_{ab}\,\left(\partial_u\,\xi^u_{(0)}- v^c\,\partial_c\,\xi^u_{(0)}\right) \label{deltag}\\
	\delta v_a &= \xi^u_{(0)}\,\partial_uv_a+\mathcal{L}_{\xi^c_{(0)}}\,v_a+ v_{a}\,\left(2\,v^c\,\partial_c\,\xi^u_{(0)}-\partial_u\,\xi^u_{(0)}\right)+ g^{(0)}_{ab}\,\partial_u \xi^b_{(0)} \nonumber \\
	&-\frac{1}{l^2} \partial_a \xi^u_{(0)}\,g^{(0)}_{uu}+ \partial_a\,\xi^u_{(0)}\,v^b\,v_b 
\end{align}
Here $\mathcal{L}_{\xi^c_{(0)}}\,g^{(0)}_{ab}$ is the Lie derivative of boundary metric $g^{(0)}_{ab}$ under boundary vector field $\xi^{(0)}_{c}$. Demanding that the transformation of free data at the boundary $\{g^{(0)}_{ab}\,,v_{a}\}$ under (\ref{bdryvec}) should obey the conditions obtained by solving variational principle will give us constraints on these boundary vector fields. The conditions that we will impose are,
\begin{align}
	\delta g^{(0)}_{\bar{z} \bar{z}}=0\,,~ \delta g^{(0)}_{z\bar{z}}=0\,,~ \delta v_a =0
\end{align}
along with $g^{(0)}_{\bar{z}\bar{z}}= v_a=0, ~ g^{(0)}_{z\bar{z}}= \frac{1}{2}$ .
From $\delta g^{(0)}_{\bar{z} \bar{z}}=0$ one gets,
\begin{align}
	\xi^z_{(0)}(u,z,\bar{z})= Y^z(u,z). \label{Yz}
\end{align}
From $\delta v_a=0$ one gets the equation,
\begin{align}
	\partial_a \xi^u_{(0)}\,g^{(0)}_{uu}= -g^{(0)}_{ab}\,\partial_u \xi^b_{(0)}. \label{deltav}
\end{align}
The $\bar{z}$ component of equation (\ref{deltav}) can be solved to give ,
\begin{align}
	\xi^u_{(0)}= f (u,z)+ \frac{l^2}{2}\,\bar{z}\,\partial_u Y^z (u,z) .
\end{align}
Imposing $ \delta g^{(0)}_{z\bar{z}}=0$ allows one to write $\xi^{\bar{z}}$ as follows,
\begin{align}
	\xi^{\bar{z}}_{(0)}= Y^{\bar{z}}(u,z)+ \bar{z} \,(2\,\partial_u f - \partial_z Y^z)+\frac{l^2}{2} \bar{z}^2\,\partial_u^2\,Y^z.
\end{align}
Also, we have $
\xi^r_{(0)}=-\partial_u \xi^u_{(0)}$. Note the explicit $\bar{z}$ dependence of the boundary vector fields.

Using these expressions one finds that the $z$ component of (\ref{deltav}) is a polynomial in $\bar{z}$ with degree $2$. Setting the coefficient of each power of $\bar{z}$ in $\delta v_z$ to zero separately one gets the following conditions,
\begin{align}
	\frac{l^2}{4}\,\partial_u^3\,Y^z+ J_1(z)\,\partial_u Y^z(u,z)&=0, \label{f} \\
	\partial^2_u\,f(u,z)+ J_{0}(z)\,\partial_u Y^z(u,z)- \partial_u\,\partial_z\,Y^z(u,z)\nonumber \\+ \frac{1}{l^2}\,\big(u^2\,\partial\,J_{1}(z)\,\partial_u\,Y^z(u,z)- u\,\sqrt{G}\,G_{1/2}(z)\,\partial_u Y^z\big)&=0 ,\label{xiz} \\
	\frac{1}{2}\,\partial_u\,Y^{\bar{z}}(u,z)-\frac{1}{l^2}\,\partial_z\,f(u,z)+\Gamma (u,z)\,\partial_u Y^z(u,z)&= 0 . \label{xibz}
\end{align}
where $\Gamma(u,z)$ was defined in equation (\ref{gamma}). After completely determining $\bar{z}$ dependence of these vector fields one can solve these differential conditions to obtain the explicit $u$ dependence.
The solution to equation (\ref{f}) is given by,
\begin{align}
	&Y^{z}(u,z)= Y(z)+\frac{\sqrt{G}\,P_{1/2}(z)}{l}\,\frac{\sin\,\left(\frac{2\,u\,\sqrt{J_1(z)}}{l}\,\right)}{\sqrt{J_1(z)}} + Y_{1}(z)\,\frac{\sin^2\left(\frac{u\,\sqrt{J_1(z)}}{l}\right)}{J_{1}(z)}.\label{Yz}
\end{align}
As the order of the differential equation is three, the solution has three undetermined integration functions $\{Y(z), P_{1/2}(z), Y_{1}(z)\}$. Similarly to obtain $f(u,z)$ one solves equation (\ref{xiz}) to find
\begin{equation}
	\begin{split}
		&f(u,z)= \sqrt{G}\,P_{-1/2}(z)+ \frac{\sin\left(\frac{2\,u\,\sqrt{J_1(z)}}{l}\right)}{\sqrt{J_1(z)}}\,L_1(u,z)+ \frac{\cos\left(\frac{2\,u\,\sqrt{J_1(z)}}{l}\right)}{J_1(z)}\,L_2(u,z) \\&+ \frac{\sqrt{G}}{4\,J_1(z)^2}\,\left(2\,J_1(z)\,(\partial P_{1/2}(z)-J_0(z)\,P_{1/2}(z))+ G_{1/2}\,Y_{1}(z)+P_{1/2}\,\partial J_1 (z)\right)  \\&+ \frac{u}{4\,J_1(z)^2}\,\left(2\,J_1(z)\,\left(\partial Y_{1}(z)- \frac{G}{l^2}\,G_{1/2}(z)P_{1/2}(z)-J_{0}(z)\,Y_{1}(z)\right)-Y_1(z)\,\partial J_{1}(z)\right) \\  &+ \frac{u}{2}\,\left(\partial Y(z)+ Y_0(z)\right) \label{fu}
	\end{split}
\end{equation}
where $L_{1}(u,z)\,,L_{2}(u,z)$ are, 
\begin{align}
	&L_{1}(u,z)=\frac{G}{4\,l\,J_{1}(z)}\,\left[2\,G_{1/2}(z)\,P_{1/2}(z)+\frac{l^2}{G}\,\left(J_{0}(z)\,Y_{1}(z)-\partial Y_{1}(z)\right)\right] \nonumber \\
	&-\frac{\sqrt{G}\,u}{4\,l\,J_{1}(z)}\,\left[G_{1/2}(z)Y_{1}(z)+2\,P_{1/2}(z)\,\partial J_{1}(z)\right]+ \frac{u^2}{4\,l\,J_{1}(z)}\,Y_{1}(z)\,\partial J_1(z),\\
	&L_{2}(u,z)=\frac{\sqrt{G}}{4\,J_{1}(z)}\,\left[2\,J_1(z)\,J_{0}(z)\,P_{1/2}(z)-G_{1/2}(z)\,Y_{1}(z)-P_{1/2}(z)\,\partial J_{1}(z)-\frac{1}{2}\,\partial P_{1/2}(z)\right]\nonumber \\
	&+\frac{u}{4\,l^2\,J_{1}(z)}\,\left[l^2\,Y_{1}(z)\,\partial J_{1}(z)-2\,G\,J_{1}(z)\,G_{1/2}(z)\,P_{1/2}(z)\right]+ \frac{\sqrt{G}\,u^2}{2\,l^2}\,P_{1/2}(z)\,\partial J_{1}(z).
\end{align}
The solution has two undetermined integration functions $P_{-1/2}(z)$ and $Y_0(z)$. Now that we know  $Y^z(u,z)$ and $f(u,z)$ in closed form, one can use them in equation (\ref{xibz}) to obtain $Y^{\bar{z}}$ as follows,
\begin{align}
	& Y^{\bar{z}}(u,z)= Y_{-1}(z)+\frac{2}{l^2}\int du\,\partial_z\,f(u,z)- 2\,\int du\,\left(\Gamma(u,z)+ J_{-1}(z)\right)\,\partial_u Y^z(u,z)+ L_3(z)
\end{align}
where $L_3(z)$ is given by,
\begin{multline}
	L_3(z)=\frac{\sqrt{G}\,P_{1/2}(z)}{2\,l\,J_{1}(z)}\,K_{1}(z)+ \frac{3\,Y_{1}(z)}{8\,J_{1}(z)^2}\,K_{2}(z)+ \frac{\partial Y_{1}(z)}{4\,J_{1}(z)^3}\,\left(J_{1}(z)\,J_{0}(z)+ \partial J_1(z)\right)\\+\frac{G\,P_{1/2}(z)}{2\,l^2\,J_1(z)^3}\left[J_{1}(z)^2\,G_{-1/2}(z)+\frac{3}{4}\,G_{1/2}(z)\,\partial J_{1}(z)-\frac{3}{2}\,G_{1/2}(z)\,J_{1}(z)\,J_0(z)\right]\\+ \frac{3\,Y_{1}(z)}{8\,J_1(z)^2}\,\bigg[\frac{2}{3}\,\partial J_{0}(z)+\frac{J_0(z)\,\partial J_1(z)}{3\,J_1(z)}-\frac{3\,\partial J_1(z)^2}{2\,J_1(z)^2}+\frac{\partial^2\,J_1(z)}{J_1(z)}-\frac{8}{3}\,J_{-1}(z)\,J_1(z)\\+\frac{5\,G\,G_{1/2}(z)}{6\,l^2\,J_1(z)}\bigg]
	+	\frac{1}{4\,J_1(z)^2}\,\left(\frac{3G}{l^2}\,G_{1/2}(z)\,\partial P_{1/2}(z)-\partial^2\,Y_{1}(z)\right)
\end{multline}
where $Y_{-1}(z)$ is another integration function and the integral over $u$ can be performed easily to obtain the closed form. $L_3(z)$ is written in a way such that the flat space limit of this vector field agrees with the locally flat solutions in \cite{Gupta:2021cwo}.

In conclusion, the boundary vector fields that generate residual diffeomorphisms on the chiral $\text{lAdS}_4$ solutions are parameterised by six holomorphic functions
\begin{align}
	\{Y_{a}(z)\,,Y(z)\,,P_{s}(z)\,\}~~{\rm for}~~ a\in \{0, \pm 1\} ~~{\rm and}~~ s \in \{\pm 1/2 \}.\label{bdry diff}
\end{align}
The subleading terms in $\frac{1}{l^2}$ as well as in $r$ expansion are given in terms of these six holomorphic functions. The solution to equations (\ref{f}-\ref{xibz}) is constructed in such a way that it has a well defined flat space limit and the behaviour is smooth if $J_1(z)$ is taken to be zero. Also one needs to take care of the length dimension of vector fields. $\xi^z$ and $\xi^{\bar{z}}$ is dimensionless but $\xi^u$ has dimension one. In the flat space limit $\xi^u$ gets scaled by $\sqrt{G}$ like we did for $g^{(1)}_{ab}$. After taking care of these subtleties one finds that in the large $l$ limit the components of boundary vector fields behave as follows,
\begin{align}
	&\xi^u_{(0)}= \sqrt{G}\,\big(P_{-1/2}(z)+ \bar{z}\,P_{1/2}(z)\big)+ \frac{u}{2}\,\big(Y_{(0)}(z)+ \partial Y(z)+ 2\,\bar{z}\,Y_1(z)\big)+ \cdots\\
	&\xi^z_{(0)}= Y(z)+ \cdots\\
	&\xi^{\bar{z}}_{(0)}= Y_{-1}(z)+\bar{z}\,Y_{0}(z)+\bar{z}^2\,Y_{1}(z)+ \cdots	.
\end{align}
where $\cdots$ denotes higher order terms in $\frac{1}{l^2}$.
One observes that in this limit the leading order terms in $\frac{1}{l^2}$ expansion of the boundary vector field  is equal to the boundary vector field that was found in the context of locally flat solutions in \cite{Gupta:2021cwo} with appropriate identification of $P_s(z)$.
If we now inspect $\delta g^{(0)}_{zz}$ then we can deduce the various transformation of background fields (\ref{funcs}) in terms of (\ref{bdry diff}). We do the following redefinition on the fields: $$ J_{1}(z) \rightarrow \frac{1}{2}\,J_{1}(z)\,,~~J_{-1}(z) \rightarrow \frac{1}{2}\,J_{-1}(z)\,~~ Y_{1}(z) \rightarrow \frac{1}{2}\,Y_{1}(z)\,~~Y_{-1}(z) \rightarrow \frac{1}{2}\,Y_{-1}(z).$$ These variations are then given by,
\begin{align}
	\delta J_a(z)&=\frac{2\,G}{l^2}\,(\tau_{a})^{ss'}\,G_s(z)\,P_{s'}(z)+\partial Y_a(z)+f_{a}^{~bc}\,J_b(z)\,Y_c(z) + \partial (J_a(z)Y(z)) \label{transJ}\\
	\nonumber \\
	\delta \kappa(z)&=Y(z)\,\partial\,\kappa(z)+2\kappa(z)\,\partial Y(z)-\partial^3\,Y(z)\nonumber \\&+\frac{2\,G}{l^2}\,(\tau^{a})^{ ss'}\,J_a(z)\,G_s(z)\,P_{s'}(z)
	+\frac{3G}{l^2}\,\epsilon^{ss'}\,G_s\,\partial P_{s'} +\frac{G}{l^2}\,\epsilon^{ss'}\,\partial G_s\,P_{s'} \\
	\nonumber \\
	\delta G_s(z)&=\frac{3}{2}\,G_s(z)\,\partial Y(z)+Y(z)\,\partial G_s(z)-2\,\partial^2\,P_s(z)+P_s(z) \,\bigg(\kappa(z)+\frac{1}{2}\,\eta^{ab}\,J_a(z)\,J_b(z)\bigg)\nonumber \\ &+\,G_{s'}(z)\,(\tau^a)^{s'}_{~s}\,Y _a(z)+4\,\partial P_{s'}(z)\,(\tau^a)^{s'}_{~s}\,J_a(z)+2\,P_{s'}(z)\,(\tau^a)^{s'}_{~s}\,\partial J_a(z)
\end{align}
Here $\{a,b,c..\} \in \{-1,0,1\}$ and $\{s,s'\} \in \{-\frac{1}{2},\frac{1}{2}\}$ and $f_{a}^{~bc}$ and $(\tau^a)^r_{~s}$ are structure constants defined as follows,
\begin{align}
	f_{ab}^{~~c}= (a-b)\,\delta^{c}_{a+b}\,,~~ (\tau _a)^r_{~s}= \frac{1}{2}(a-2s)\,\delta^{r}_{a+s} \label{structure_const}
\end{align}	
The indices $s,s'$ are raised and lowered by $\epsilon ^{ss'}$ and $\epsilon_{ss'}$ whereas indices  $a,b,c,..$ are raised and lowered by $\eta^{ab}$ and $\eta_{ab}$. These are further defined as,
\begin{align}
	\eta_{ab}=-(1-3a^2)\,\delta_{a+b,0},~~ \eta ^{ab}= - \frac{1}{(1-3a^2)}\,\delta^{a+b,0} \\
	\epsilon_{ss'}= 2s'\,\delta_{s+s',0},~~\epsilon^{ss'}= -2s'\,\delta^{s+s',0} \label{killingform}
\end{align} 
\subsection{Algebra of vector fields}
In the previous section we derived full set of vector fields that enable us to move in the class of $\text{lAdS}_4$ solutions. Now we check whether the Lie brackets between any two of these vector fields close under the modified commutator (see for instance \cite{Barnich:2010eb})
\begin{align}
	[\,\xi_1\,,\,\xi_2\,]_M :=\,\xi_1^{\sigma}\,\partial_{\sigma}\,\xi_2-\,\xi_2^{\sigma}\,\partial_{\sigma}\,\xi_1-\,\delta_{\xi_1}\,\xi_2+\,\delta_{\xi_2}\,\xi_1 .\label{lie brac}
\end{align}
Calculating the bracket as in eq. (\ref{lie brac}) we find:
\begin{align}
	&[\,\xi_1(P^1_r,Y^1,Y^1_{a})\,,\,\xi_2(P^2_r,Y^2,Y^2_{a})\,]_M=\hat{\xi}^u_{(0)}\,\partial_u+\hat{\xi}^z_{(0)}\,\partial_z+\hat{\xi}^{\bar{z}}_{(0)}\,\partial_{\bar{z}}+ \hat{\xi}^r_{(0)}\partial_r+ \cdots \label{lie brac 2}
\end{align}
where
$\{\hat{\xi}^u_{(0)}\,, \hat{\xi}^z_{(0)}\,, \hat{\xi}^{\bar{z}}_{(0)}\,, \hat{\xi}^r_{(0)}\}$ are functions of $\hat{P}_r\,, \hat{Y}\,,\hat{Y}_a$ which are defined as follows,
\begin{equation}
	\begin{split}
		\hat{Y}&= Y^{1}(z)\,\partial Y^{2}(z)-Y^{2}(z)\,\partial Y^{1}(z)- \frac{2\,G}{l^2}\,\epsilon^{rs}\,P^{1}_{r}(z)\,P^{2}_{s}(z), \\
		\hat{Y}_a&=Y^1(z)\,\partial Y^{2}_a-Y^2(z)\,\partial Y^{1}_a -f_{a}^{~bc}\,Y^1_{b}(z)Y^2_{c}(z)-\frac{4\,G}{l^2}\,(\tau_a)^{rs}\,P^1_{r}\,\partial P^{2}_{s} \\
		&+\frac{4\,G}{l^2}\,(\tau_a)^{rs}\,P^2_{r}\,\partial P^{1}_{s}+\frac{4\,G}{l^2}\,J_a(z)\,\epsilon^{rs}P^{1}_r\,P^2_s, \\
		\hat{P}_r&=-(\tau^a)^s_{~r}\,\left(P^{1}_s\,Y^2_a-P^{2}_s\,Y^1_a\right) + Y^1\,\partial P^2_{r}-  Y^2\,\partial P^1_{r}+ \frac{1}{2}\,P^1_{r}\,\partial Y^2+ \frac{1}{2}\,P^2_{r}\,\partial Y^1. \label{resPr}
	\end{split}
\end{equation}
The vector fields on L.H.S of (\ref{lie brac}) are parameterised by two sets of six holomorphic functions written down in (\ref{bdry diff}) and under this modified commutator the form of resultant vector field in R.H.S does not change and have the same functional dependence on combination of these two sets as written down in (\ref{resPr}). Note the dependence of $\hat{Y}^a$ on the background function $J_a$. Using (\ref{resPr}) one can write down the commutators between various modes of (\ref{bdry diff}) as follows, 
\begin{itemize}
	\item Denoting the parameter $Y_a(z) = - z^n$  by ${\cal J}^a_{n}$ for $a \in \{ -1, 0, 1\}$ and $n \in {\mathbb Z}$. 
	%
	The commutator of these is given by,
	\begin{align}
		\label{sl2alg1}
		[{\cal J}^{a}_m\,,\, { \cal J}^{b}_n\,]=\big(a-b \big)\,{\cal J}^{a+b}_{m+n}\
	\end{align}
	This can be recognised as the $\mathfrak{sl}_2$ current algebra. 
	\item Denoting the parameter $Y(z)$ as $Y(z) = -z^{n+1}$ by ${\cal L}_n$ for $n\in {\mathbb Z}$ we find: 
	\begin{align}
		\label{sl2alg2}
		[{\cal L}_m, {\cal L}_n] = (m-n) \, {\cal L}_{m+n}, ~~~ [\,{\cal L}_m\,,{\cal J}^a_{n}\,]=-n\,{\cal J}^{a}_{m+n}
	\end{align} 
which are Witt algebra, with $\mathcal J^a(z)$ being $h=1$ currents.
	\item Finally denoting the parameter $P_s(z)$ as $\mathcal{P}_{s,r}(z)=-z^{r+\frac{1}{2}}$
	where $r \in {\mathbb Z}+\frac{1}{2}$ and $s\in \{ - \frac{1}{2}, \frac{1}{2}\}$, the remaining commutators work out to be:
	\begin{align}
		\label{sl2alg3}
		[{\cal L}_n, {\cal P}_{s,r}] = \frac{1}{2} (n-2 \, r) \, {\cal P}_{s, n+r}, ~~ [{\cal J}^a_{n}, {\cal P}_{s,r}] = \frac{1}{2} (a-2 \, s) \, {\cal P}_{a+s, n+r} =0 
	\end{align}
	%
	\begin{align}
		\big[\mathcal{P}_{s,r}, \mathcal{P}_{w,t} \big]= \frac{2G}{l^2} (r-t)\,\mathcal{J}_{s+w,t+r}-\frac{2G}{l^2}\epsilon_{s w}\, \mathcal{L}_{r+t} +\frac{4G}{l^2}\epsilon_{sw} \,\left(\mathcal{J}^{a}_{t+r+1} J_{a}(z)\right)\label{sl2alg4}
	\end{align}
\end{itemize}
Thus the final result of the algebra of the vector fields that preserve our space of locally AdS$_4$ solutions are (\ref{sl2alg1} -  \ref{sl2alg4}). This algebra is one of the results of this paper. The symmetry algebra can be considered as the chiral version of $\Lambda$-$\mathfrak{bms}_4$ \cite{Compere:2019bua} and is a Lie algebroid as indicated by the presence of background field $(J_a(z))$ dependent structure constants in the commutator of two $\mathcal{P}_{r,s}$. We denote this algebra as chiral $\Lambda$-$\mathfrak{bms}_4$ algebra. In the limit $l\rightarrow \infty$ one can see that $\mathcal{P}_{s,r}$ commutes and resultant algebra becomes chiral $\mathfrak{bms}_4$ (\ref{comm}).
\subsection{Charges}
If we perform the following redefinition of the field $\kappa(z)$ and introduce a new function $T(z)$ as follows,
$$T= \kappa(z)-\frac{1}{2}\,\eta^{ab}\,J_{a}(z)\,J_{b}(z)$$
then the resultant transformations of the field $T(z)$ and $G_{s}(z)$ will be,
\begin{align}
	\label{modified T}
	\delta T(z)&=Y(z)\,\partial\,\kappa(z)+2\kappa(z)\,\partial Y(z)-\partial^3\,Y(z)\nonumber \\&-\eta^{ab}\,J_{a}(z)\,\partial Y_b(z)
	+\frac{3G}{l^2}\,\epsilon^{ss'}\,G_s\,\partial P_{s'} +\frac{G}{l^2}\,\epsilon^{ss'}\,\partial G_s\,P_{s'},\\
	\nonumber\\
	\delta G_s(z)&=\frac{3}{2}\,G_s(z)\,\partial Y(z)+Y(z)\,\partial G_s(z)-2\,\partial^2\,P_s(z)+P_s(z) \,\bigg(T(z)+\eta^{ab}\,J_a(z)\,J_b(z)\bigg)\nonumber \\ &+\,G_{s'}(z)\,(\tau^a)^{s'}_{~s}\,Y _a(z)+4\,\partial P_{s'}(z)\,(\tau^a)^{s'}_{~s}\,J_a(z)+2\,P_{s'}(z)\,(\tau^a)^{s'}_{~s}\,\partial J_a(z). \label{modified G}
\end{align}
The variation of holomorphic functions $\{T(z), J_{a}(z), G_{s}(z)\}$ in (\ref{transJ}, \ref{modified T}, \ref{modified G}) is very similar to the transformation of chiral currents in $2d$ CFT. For instance under $Y(z)$, $T(z)$ transforms like a conformal stress tensor of of weight $2$ of a $2d$ CFT.  Therefore, we treat it as a stress tensor and $Y(z)$ as generator of Virasoro transformations. Then under these Virasoro transformations $J_a(z)$ and $G_s(z)$ transform as conformal primaries of weight $1$ and $\frac{3}{2}$ respectively. The transformation of $J_a(z)$ under $Y_a$ and the form of structure constant $(f_{a}^{~bc})$ suggest that these are $\mathfrak{sl}(2,\mathbb{R})$ Kac Moody currents at some (undetermined) level and therefore the parameters $Y_a(z)$ generate this current algebra symmetry. The transformation of $G_s(z)$ under this current algebra indicates that these $G_s(z)$ are doublet of current algebra primaries. To summarize, the background fields can be classified as a chiral stress tensor $T(z)$, a triplet of $\mathfrak{sl}(2,\mathbb{R})$ Kac Moody currents $J_{a}(z)$ and a doublet $(h=\frac{3}{2})$ of conformal primary as well as current algebra primaries $G_{s}(z)$. 

We now promote the fields in (\ref{funcs}) to conformal operators and propose a line integral charge just like in a $2d$ CFT that will induce the transformations  (\ref{transJ}, \ref{modified T}, \ref{modified G}) of background fields. Once we propose such a charge, we derive the OPEs between various operators  in (\ref{funcs}) that will give rise to these transformations. The conjectured line integral charge is,
	\begin{align}
		Q= \oint\,\frac{dz}{2\pi i} \,\left[T(z)\,Y(z)- \eta^{ab}\,J_{a}(z)Y_b(z)+ \frac{2G}{l^2}\epsilon^{ss'}\,G_{s}(z)\,P_{s'}(z)\right]\label{charge}
	\end{align}
Here we assume that the underlying symmetry algebra is at a fixed values of central charge and current level.\footnote{It still remains to be seen how such a charge may be derived from the bulk using first principles.} The operator product expansions between various operators in (\ref{funcs}) are given by
\begin{align}
	 T(z)\,J_a(w)&= \frac{J_a(w)}{(z-w)^2}+ \frac{\partial J_a(w)}{(z-w)} ,\cr
	J_a(z)\,J_b(w) &= -\frac{\,\eta_{ab}}{(z-w)^2}- \frac{f_{ab}^c\,J_c(w)}{(z-w)} ,\cr
	T(z)\,T(w) &= -\frac{6}{(z-w)^4}+ \frac{2 T(w)}{(z-w)^2}+ \frac{\partial T(w)}{(z-w)},\cr
	J_a(z)\,G_s(w) &= -\frac{G_{s'}\,(\tau_a)^{s'}_{s}}{(z-w)}, \cr
	 T(z)\,G_s(w) &= \frac{3\,G_s(w)}{2\,(z-w)^2}+ \frac{\partial G_s(w)}{(z-w)},\cr
	\frac{2G}{l^2}\,G_s(z)\,G_{s'}(w) &= \frac{4\,\epsilon_{ss'}}{(z-w)^3}- \frac{\epsilon_{ss'}\,T(w)}{(z-w)}- \frac{\epsilon_{ss'}\, J^2(w)}{(z-w)} \nonumber \\ &- 2\,(\tau)^a_{~ss'}\,\bigg[\frac{\partial J_a (w)}{(z-w)}+ \frac{2\,J_a(w)}{(z-w)^2}\bigg]. \label{OPE_relation}
\end{align}
These OPEs are obtained using standard $2d$ conformal field theory techniques. Next we write the operators $T(z)$, $J_a(z)$ and $G_s(z)$ in terms of their modes as
\bea
L_n = \oint \frac{dz}{2\pi i} z^{n+1} T(z), ~~ J_{a,n} = \oint \frac{dz}{2\pi i} z^{n} J_a(z), ~~ G_{s,r} = \oint \frac{dz}{2\pi i} z^{r+ \frac{1}{2}} G_s(z)
\eea
The OPEs \eqref{OPE_relation} are then translated  to the following commutators,
	\begin{align}
	\label{bulk_comm}
	&[L_m, L_n] = (m-n) \, L_{m+n} - \frac{1}{2} m (m^2-1) \delta_{m+n,0} \nonumber \\
	&[L_m, J_{a,n}] = -n \, J_{a, m+n}, ~~~ [L_m, G_{s,r}] = \frac{1}{2}(n-2r) \, G_{s,n+r} \nonumber\\
	&[J_{a,m}, J_{b,n}] = - m \, \eta_{ab} \, \delta_{m+n, 0} - {f_{ab}}^c \, J_{c,m+n}\,,~~[J_{a,n}\,, G_{s,r}] = -G_{s',n+r} {(\lambda_a)^{s'}}_s\nonumber \\
	& \!\!\!\!\!\frac{2G}{l^2} [G_{s,r}, \! G_{s',r'}] \!= \!\epsilon_{ss'} \!\! \left[2 \! \left(r^2 - \frac{1}{4} \right) \! \delta_{r+r',0} \! - \!  \,  L_{r+r'}  \! - \!  \, (J^2)_{r+r'} \right] \!\!- \! (r-r')  J_{a, r+r'}  {(\lambda^a)}_{ss'}
	\end{align}
where we shifted the zero mode of $L_{n}$ as,
\begin{align}
	L_{n} \rightarrow L_{n}+\frac{1}{2}\delta _{n,0}
\end{align}
The commutator algebra in (\ref{bulk_comm}) is the main result of the paper. Under a suitable scaling of operator $G_{s} \rightarrow \frac{l^2}{2G}\,G_{s}$ and taking $l \rightarrow \infty$, the algebra (\ref{bulk_comm}) reduces to chiral $\mathfrak{bms}_4$ with non-zero central charge and current level.
\paragraph{Comments on the charges}: It is well known that the usual ways of calculating charges through covariant phase space formalism \cite{Wald:1999wa, Iyer:1994ys, Lee:1990nz} or the cohomological formalism  \cite{Barnich:2000zw, Barnich:2001jy, Barnich:2007bf} lead to co-dimension two integral charges. It can be shown that for our chiral locally AdS$_4$ solutions, exploiting the following ambiguity in surface charge two-forms $\mathbf{k}_{\xi}$  \cite{Compere:2016jwb}
\begin{align}
	\mathbf{k}_{\xi} \rightarrow \mathbf{k}_{\xi}+ d \mathbf I_{\xi}
\end{align}
 the co-dimension two charges vanishes, where  $\mathbf{I}_{\xi}$ is a three-form which is not fixed by the theory. Therefore only charges that could exist for these solutions are the co-dimension three charges (line integral from bulk 4d perspective) defined in \eqref{charge}. This conclusion is in agreement with the results of  \cite{Compere:2008us}, where the authors showed that for the Neumann boundary condition, the effective boundary theory is an induced gravity theory for which it is expected that such codimension-two charges will vanish since the boundary diffeomorphisms are pure gauge transformations. If one is able to derive this induced gravity action, one can in principle  calculate line integral charges from such a $3d$ action.
 Therefore, the line integral charges that we have postulated in \eqref{charge} ought to be the feature of Neumann boundary conditions where the boundary theory is gravitating such that these are co-dimension two charges from the perspective of induced gravity theory at the boundary(co-dimension three charges from $4d$ bulk perspective). Finally, even though the co-dimension two charges vanish for these locally AdS$_4$ solutions, it does not mean that the asymptotic symmetries uncovered in this work are pure gauge transformations as the flat space limit of these chiral locally AdS$_4$ solutions have physical interpretation as the degenerate vacua of $\mathbb R^{1,3}$ gravity associated with the spontaneous symmetry breaking of chiral $\mathfrak{bms}_4$ algebra with the boundary currents \eqref{currents} identified as the Goldstone modes  \cite{Gupta:2021cwo}. We hope to substantiate some of these expectations in our future works.

 We have derived a charge algebra from the bulk AdS$_4$ gravity after postulating the existence of line integral charge that generates the residual diffeomorphisms which themselves obey the algebra (\ref{sl2alg1}, \ref{sl2alg2}, \ref{sl2alg3}, \ref{sl2alg4}). This charge algebra is non- linear as can be seen by the presence of $(J^2)_{r+r'}$ term on the R.H.S of commutator $\left[G_{s,r}, G_{s',r'}\right]$ and is a semi-classical limit of a $\mathcal{W}$-algebra at a specific value of central charge and current level. The contents of this $\mathcal{W}$-algebra include,
\begin{enumerate}
	\item  A chiral stress tensor $T(z)$ with a fixed central charge $c$.
	\item  A triplet $J_a$ for $a \in \{-1,0,1\}$ satisfying  $\mathfrak{sl}(2,\mathbb{R})$ current algebra with level $\kappa$.
	\item A doublet of spin-$\frac{3}{2}$ chiral primary operators $G_s(z)$ for $s \in \{-\frac{1}{2},\frac{1}{2}\}$.
\end{enumerate} 
To validate our findings, in next section we will derive a complete quantum version of the algebra (\ref{bulk_comm}) for arbitrary central charge $c$ and current level $\kappa$ using the standard tools of $2d$ CFT such as associativity constraints and show that with the correct identification of the operators  this quantum algebra in the semi-classical limit is identical to \eqref{bulk_comm}. We will also show that there exists only two possible algebras with the same operator content, one is chiral $\mathfrak{bms}_4$ algebra \eqref{comm} with non zero central charge and other is the quantum version of (\ref{bulk_comm}).
\section{Derivation of the $\mathcal{W}$-algebra from $2d$ CFT }
\label{sec2}
The OPEs of primary operators in $2d$ chiral CFT are demanded to be consistent with conformal symmetry as well as satisfy the associativity constraints \cite{Belavin:1984vu}. One can implement the associativity constraints in terms of Jacobi Identities of the modes of primary operators using the procedure laid out fully by Nahm et al. \cite{Blumenhagen:1990jv} (see also \cite{Blumenhagen:2009zz}) developing further on earlier works such as \cite{Bowcock:1990ku}. Even though these techniques of construction of new algebras are standard but for the convenience of the readers we review them and show their working explicitly to derive the quantum version of $\mathcal W$-algebra \eqref{bulk_comm}.

 First we list all the simple Virasoro primary fields in our case.\footnote{A primary is said to be simple if the corresponding state is orthogonal (with overlap defined using the BPS duals) to those of all the normal ordered quasi-primaries \cite{Blumenhagen:1990jv}.}  Identity operator ${\mathbb I} (h=0)$,  a chiral stress tensor $T(z) (h=2)$,  $\mathfrak{sl}(2, {\mathbb R})$ currents} $J_a(z) (h=1)$, the current algebra primary doublet $G_s(z) (h=3/2)$. The conformal dimensions of these simple primaries were fixed by the calculations of the previous section.
 Then one constructs all the quasi-primaries that are formed by normal ordered products of simple primaries, that can appear on the right hand side of the OPE of simple primaries with dimensions up to $2h-1$ where $h$ is the largest conformal dimension in the set of simple primaries. In our case the highest dimension is that of $T(z)$ which has $h=2$. In this case, we will see that we only have to consider quadratic normal ordered quasi-primaries of dimension up to $2$. 
 Then one uses global conformal symmetry (and other symmetries postulated) to solve for the structure constants that appear between simple primaries and global descendants of quasi-primaries completely. However, those between three simple primaries remain free parameters at this stage. 
 Finally one needs to impose the Jacobi identities on the corresponding commutators of modes of the simple primaries. This, in principle, determines the remaining structure constants. 
The theorems of Nahm et al. \cite{Blumenhagen:1990jv} then guarantee that the resultant chiral algebra satisfies all the necessary constraints of $2d$ chiral CFTs.
\subsection*{Normal ordered quasi-primaries of interest}
The relevant quasi-primaries that are normal ordered product between any two of the simple primaries defined in the last paragraphs are,
{\small
	$$(TT)(w),~ (TJ_a)(w), ~(TG_s)(w),~ (J_aJ_b)(w), ~(J_aG_s)(w), ~ (G_s G_{s'})(w), (J_aJ_bJ_c)(w), ~ \cdots $$ }
and their derivatives. Among these there is exactly one chiral operator of dimension $h \le 2$ and it is $(J_aJ_b)(w)$ which has $h=2$. It is easy to see that the symmetric tensor $(J_{(a}J_{b)})(w)$ is a quasi-primary (whereas the anti-symmetric combination is not). This quasi-primary belongs to a reducible representation of the $\mathfrak{sl}(2, {\mathbb R})$ algebra generated by the zero modes of $J_a(z)$; which can be decomposed into irreps by separating the trace (constructed using the Killing form $\eta_{ab}$ of $\mathfrak{sl}(2, {\mathbb R})$) and the traceless parts. So the only normal ordered quasi-primaries we need to consider are: 
$$(J^2)(w) := \eta^{ab} (J_aJ_b)(w) ~~{\rm and} ~~ (J_{(a}J_{b)})(w) - \frac{1}{3} \eta_{ab} (J^2)(w).$$
Note that these are a singlet ($j=0$) and a quintuplet ($j=2$) respectively of the $\mathfrak{sl}(2, {\mathbb R})$ algebra. 
\subsection{The OPE Ansatz}
\label{sec3}
To construct the algebra that we seek, we  list out OPEs that are unambiguously fixed already by the nature of the chiral operators we have postulated, namely, one $T(z)$ ($h=2$), one $J_a(z)$ ($h=1$) and one $G_s(z)$ ($h=3/2$). These are:
\bea
\label{ope-one}
 T(z) T(w) &\sim& \frac{c}{2} (z-w)^{-4} + 2 \,  (z-w)^{-2} \, T(w) +  (z-w)^{-1} \, \partial T(w)  \, ,\cr
 && \cr
 T(z) J_a(w) &\sim& (z-w)^{-2} J_a(w) + (z-w)^{-1} \partial J_a(w) \, ,\cr
 && \cr
T(z) G_s(w) &\sim& (3/2) (z-w)^{-2} G_s(w) + (z-w)^{-1} \partial G_s(w)  
\eea
and, following the conventions of Polyakov in \cite{Polyakov:1987zb} for the $\mathfrak{sl}(2, {\mathbb R})$ current algebra 
\bea
\label{ope-two}
J_a(z) J_b(w) &\sim& -\frac{\kappa}{2} \eta_{ab} (z-w)^{-2} + (z-w)^{-1}{f_{ab}}^c J_c(w), \label{JJ OPE} \cr
&& \cr
J_a(z) G_s(w) &\sim& (z-w)^{-1} G_{s'}(w) {(\lambda_a)^{s'}}_s 
 \eea
where $a,b, \cdots = 0, \pm 1$ and $s, s', \cdots = \pm 1/2$. The matrices $\lambda_a$ form a $2\times 2$ matrix representation of $\mathfrak{sl}(2, {\mathbb R})$ algebra which we take to be:
\bea
[\lambda_a, \lambda_b] = (a-b) \, \lambda_{a+b} \, .
\eea
We use the following explicit form of these matrices
 %
\bea
\label{lambdaudds}
{(\lambda_a)^s}_{s'} = \frac{1}{2} (a-2s') \delta^{s}_{a+s'}  ~ .
\eea
Note that the definition of $(\lambda_a)^s_{~s'}$ is equal to the structure constant $(\tau_a)^s_{~s'}$ determined from  the bulk AdS$_4$ gravity calculation  in \eqref{structure_const}. We define the $\mathfrak{sl} (2, {\mathbb R})$ Killing form $\eta_{ab}$ via ${\rm Tr} \left( \lambda_a \lambda_b \right) = \frac{1}{2} \eta_{ab}$. Using the $\lambda_a$ in (\ref{lambdaudds}) we have 
\bea
{\rm Te} (\lambda_a \lambda_b) = {(\lambda_a)^{s'}}_s {(\lambda_b)^s}_{s'} = \frac{1}{2} (3a^2-1) \delta_{a+b,0} 
\eea
which means $\eta_{ab} = (3a^2-1) \delta_{a+b,0}$ along with its inverse $\eta^{ab} = (3a^2-1)^{-1} \delta_{a+b,0}$.  This is identical to $\eta_{ab}$ used in \eqref{killingform}.

There is a unique invariant tensor in the tensor product of two spinor ($j=1/2$) representations of $\mathfrak{sl} (2, {\mathbb R})$, namely the antisymmetric $\epsilon_{ss'}$ and we define it such that $\epsilon_{-\frac{1}{2}, + \frac{1}{2}} = 1$. We can lower and raise the fundamental/spinor indices on ${(\lambda_a)^s}_{\tilde s}$ using $\epsilon_{s\tilde s}$ and its inverse $\epsilon^{s\tilde s}$, and the adjoint indices using $\eta_{ab}$ and its inverse $\eta^{ab}$. 
%
%
%
%
%
%
%
%
\subsection*{Ansatz for $G_s(z) G_{s'}(w)$ OPE :}
The only remaining OPE is that of $G_s(z) G_{s'}(w)$ and using global conformal invariance we can write the right hand side in terms of the quasi-primaries of dimension up to (and including) $h=2$ and their global descendants. The list of such quasi-primaries is 
\bea 
\label{allquasis}
\Big{\{} {\mathbb I}, ~ J_a(w), ~ G_s(w), ~T(w), ~(J^2)(w), ~(J_{(a}J_{b)})(w)- (1/3) \eta_{ab} (J^2)(w) \Big{\}} 
\eea
Apart from the (global) conformal symmetry this OPE has to be consistent with the $\mathfrak{sl}(2, {\mathbb R})$ symmetry (the global part of the $\mathfrak{sl}(2, {\mathbb R})$ current algebra) as well. From the OPE of $J_a(z)$ with $G_s(w)$ we know that the index $s$ is a doublet ($j=1/2$) index of this $\mathfrak{sl}(2, {\mathbb R})$. Therefore the quasi-primaries that can appear in the OPE $G_s(z) G_{s'}(w)$ can only carry $\mathfrak{sl}(2, {\mathbb R})$ indices of irreps in the tensor product of two doublets, namely the singlet $(j=0)$ and the triplet $(j=1)$. Therefore, in our list (\ref{allquasis}) of quasi-primaries neither $G_s(w)$ nor $(J_{(a}J_{b)})(w)- (1/3) \eta_{ab} (J^2)(w)$ can appear as they correspond to $j=1/2$ and $j=2$ irreps of $\mathfrak{sl}(2, {\mathbb R})$ respectively.
%
\begin{itemize}
\item The coefficients in front of the $j=0$ quasi-primary operators $\{{\mathbb I}, T(w), (J^2)(w)\}$ have to be proportional to the Clebsch-Gordan (CG) coefficients of ${\bf \frac{1}{2}} \otimes {\bf \frac{1}{2}} \rightarrow {\bf 0}$, namely $\epsilon_{ss'}$. 
\item The coefficient in front of the $j=1$ quasi-primary operator $J_a(w)$ has to be proportional to the C coefficients of ${\bf \frac{1}{2}} \otimes {\bf \frac{1}{2}} \rightarrow {\bf 1}$, namely $(\lambda^a)_{ss'}$. 
\end{itemize}
Thus after using all the global symmetries we arrive at the ansatz: 
\bea
\label{ope-three}
G_s(z) G_{s'}(w) &=&  2 \alpha \, \epsilon_{ss'} (z-w)^{-3} + \delta \, (z-w)^{-2} \, {(\lambda^a)}_{ss'} \left[ 2 \, J_a(w) + (z-w) \, {J_a}'(w) \right] \cr
&& \!\!\!\!\! + \beta \, \epsilon_{ss'} (z-w)^{-1} \,   T(w) + \gamma \, (z-w)^{-1} \, \epsilon_{ss'} (J^2)(w)  ~.
\eea
The parameters ($\alpha$, $\beta$, $\gamma$, $\delta$) in (\ref{ope-three}) as well as the central charges $(c, \kappa)$ in (\ref{ope-one}, \ref{ope-two}) are to be constrained using OPE associativity, which we implement as Jacobi identities of commutators of the modes of our simple primaries. Writing the operators $T(z)$, $J_a(z)$ and $G_r(z)$ in terms of their modes as
\bea
L_n = \oint \frac{dz}{2\pi i} z^{n+1} T(z), ~~ J_{a,n} = \oint \frac{dz}{2\pi i} z^{n} J_a(z), ~~ G_{s,r} = \oint \frac{dz}{2\pi i} z^{r+ \frac{1}{2}} G_s(z)
\eea
with $n \in {\mathbb Z}$ and $r \in {\mathbb Z}+\frac{1}{2}$.
 The OPEs (\ref{ope-one}, \ref{ope-two}, \ref{ope-three}) translate to:
\bea
\label{LL-comm}
[L_m, L_n] = (m-n) \, L_{m+n} + \frac{c}{12} m (m^2-1) \delta_{m+n,0}
\eea
\bea
\label{LJ-comm}
[L_m, J_{a,n}] = -n \, J_{a, m+n}, ~~~ [L_m, G_{s,r}] = \frac{1}{2}(n-2r) \, G_{s,n+r}
\eea
\bea
\label{JJ-comm}
[J_{a,m}, J_{b,n}] = -\frac{1}{2} \kappa \, m \, \eta_{ab} \, \delta_{m+n, 0} + {f_{ab}}^c \, J_{c,m+n}
\eea
\bea
\label{JG-comm}
[J_{a,n}, G_{s,r}] = G_{s',n+r} {(\lambda_a)^{s'}}_s
\eea
\bea
\label{GG-comm}
&& \!\!\!\!\! [G_{s,r}, \! G_{s',r'}] \!= \!\epsilon_{ss'} \!\! \left[\alpha \! \left(r^2 - \frac{1}{4} \right) \! \delta_{r+r',0} \! + \! \beta \,  L_{r+r'}  \! + \! \gamma \, (J^2)_{r+r'} \right] \!\!+ \!\delta (r-r')  J_{a, r+r'}  {(\lambda^a)}_{ss'} \cr &&
\eea
where $(J^2)_{r+r'}$Finally, following \cite{Blumenhagen:1990jv} we have to impose the Jacobi identities on these commutation relations. We turn to this next.
%
\subsection{Imposing Jacobi identities}
\label{sec4}
There are three species of modes $\{L_m, J_{a,n}, G_{s,r}\}$ and so a total of ten classes of Jacobi identities to impose. We choose to denote them schematically by $(LLL)$, $(LLJ)$, $(LJJ)$, $(JJJ)$, etc. in a self-evident notation. Imposition of $(LGG)$ gives,
\begin{align}
\alpha -\frac{c}{6} \beta +\gamma \, \frac{\kappa}{2} =0,
\end{align}
whereas from $(JGG)$, one gets
\bea
\label{finalconds}
 \alpha -\delta \, \frac{\kappa}{2} =0, ~~ \beta - \gamma \, (\kappa +2) - \frac{1}{2} \delta =0,  \label{alleqn}
\eea
The last Jacobi identity left is:
\bea
\label{ggg}
[G_{s,r},[G_{s',r'}, G_{\tilde s, \tilde r}]] + [G_{\tilde s, \tilde r}, [G_{s,r}, G_{s',r'}]]- [G_{s',r'}, [G_{s,r}, G_{\tilde s, \tilde r}]] =0 ~.
\eea
To impose this we  need the commutator $[ (J^2)_n, G_{s,r}]$ which has two different expressions, namely
{\small
	\bea
	\label{J2G-one}
	[(J^2)_n, G_{s,r}] \!=\! \eta^{ab} \!\left[ \!(J_a G_{s'})_{n+r} {(\lambda_b)^{s'}}_s + (G_{s'} J_b )_{n+r} {(\lambda_a)^{s'}}_s\right]\! +\! \frac{3}{4} \left(r + \frac{1}{2} \right) G_{s, n+r} 
	\eea
}
and 
{\small
	\bea
	\label{J2G-two}
	[(J^2)_n, G_{s,r}] \!&=&\! 2 \eta^{ab} (J_a G_{s'})_{n+r} {(\lambda_b)^{s'}}_s - \frac{3}{4} (n+1) \, G_{s, n+r}~.
	\eea
}
Taking the linear combination (\ref{J2G-two}) +2 $\times$ (\ref{J2G-one}) we arrive at:\footnote{This particular linear combination dictated by the requirement that the right hand side is written in terms of modes of the quasi-primaries $G_s(z)$ and $(J_a G_s)(z) + (1/2) (G_s J_a)(z)$.}
%
{\small
	\bea
	\label{J2G-final}
	\!\![ (J^2)_n, G_{s,r}] \!\!=\!  \eta^{ab} \left[ \frac{4}{3} \, (J_a G_{s'})_{n+r} {(\lambda_b)^{s'}}_s +\frac{2}{3} \, (G_{s'} J_a)_{n+r} {(\lambda_b)^{s'}}_s \right] \!\! - \! \frac{1}{4} (n-2r) G_{s,n+r} \, .
\eea
}

Using (\ref{J2G-final}) along with the other commutators we find that in \eqref{ggg} the terms containing $(J_a G_{\hat s})_R$, and $(G_{\hat s} J_a)_R$ do not survive and the remaining terms are proportional to $G_{s,R}$ and we find that LHS of \eqref{ggg} is given by
{\small
\bea
&&  \epsilon_{s'\tilde s} \, G_{s,R} \left[ \left(\frac{\beta}{2} -\frac{\gamma-\delta}{4} \right) (R-3r)   \right]  \cr
&& \hskip 1cm + \epsilon_{\tilde s s} \, G_{s',R} \left[ \left(\frac{\beta}{2} -\frac{\gamma-\delta}{4} \right) (R-3r')  \right]  \cr
&& \hskip 2cm + \epsilon_{s s'} \, G_{\tilde s,R} \left[ \left(\frac{\beta}{2} -\frac{\gamma-\delta}{4} \right)(R-3 \tilde r)    \right]  ~ .\nonumber
\eea
}
This can vanish, for generic values of the floating indices, only if 
\bea \beta=\frac{\gamma-\delta}{2}.\eea
We have obtained linear and homogeneous equations in the variables $(\alpha,\beta,\gamma,\delta)$,  which can be solved for $c$ (for $\kappa \ne - 5/2$) as
\bea
\label{final-result-one}
c = - \frac{6 \kappa \, (1+ 2 \kappa)}{5+2\kappa} ~ .
\eea
When this holds, for generic values of $\kappa$, there exists one non-trivial solution, which can be written (taking $\gamma$ to be the independent variable) as
\bea
\label{final-result-two}
\alpha = - \frac{1}{4} \gamma \kappa (3+2\kappa), ~~ \beta =  \frac{1}{4}\gamma(5+2\kappa), ~~ \delta = -\frac{1}{2} \gamma (3+2\kappa) \, . 
\eea
This is our final result valid for generic values of $\kappa$ ($ \ne -5/2$) . The full commutator algebra is \eqref{RESULT} which we denote by $\mathcal{W}(2;(3/2)^2,1^3)$. Some observations are in order:
\begin{itemize}
	\item As mentioned in the introduction, a $\mathcal{W}$-algebra isomorphic to that of $\mathcal{W}(2;(3/2)^2,1^3)$ already existed in the literature  \cite{Romans:1990ta}. The current algebra in \cite{Romans:1990ta} was $\mathfrak{sp}(2)$ which is isomorphic to $\mathfrak{sl}(2,\mathbb{R})$. The central charge \eqref{final-result-one} is identical for both the algebras.
\item When $c \ne - \frac{6 \kappa \, (1+ 2 \kappa)}{5+2\kappa}$ we have to set $\alpha =\beta = \gamma = \delta =0$, and the corresponding algebra is a chiral extension of $\mathfrak{iso}(1,3)$ with arbitrary $c$ and $\kappa$.
\vskip .1cm
\hskip 1cm This case corresponds to the algebra found in \cite{Banerjee:2020zlg, Banerjee:2021dlm} as the central charges there remained undetermined. 
\item We may also consider the limit $\gamma \rightarrow 0$ in the non-trivial solution (\ref{final-result-two}). This corresponds to a contraction of the algebra (\ref{GG-comm}) to the chiral extension of $\mathfrak{iso}(1,3)$, but with the relation (\ref{final-result-one}) still in place.

\hskip 1cm The one found in \cite{Gupta:2021cwo} from an analysis of chiral boundary conditions on ${\mathbb R}^{1,3}$ or ${\mathbb R}^{2,2}$ gravity appears to belong to this class.
 
\item When $\gamma \ne 0$ its value can be fixed to be any non-zero function of $\kappa$ by an appropriate normalisation of the generators $G_{s,r}$. Then the algebra is determined completely in terms of one parameter (say) $\kappa$.
\end{itemize}
\vskip .3cm 
\subsection*{Relation to $\mathfrak{so}(2,3)$}
\vskip .2cm 
Now we will argue that $\mathcal{W}(2;(3/2)^2,1^3)$ is based on (and is an extension of) $\mathfrak{so}(2,3)$. For this, first note that $\mathfrak{so}(2,3)$ algebra can be written in a more suggestive manner as follows:
\bea
[L_m, L_n] &=& (m-n) L_{m+n}, ~~ [\bar L_m, \bar L_n] = (m-n) \, \bar L_{m+n}, ~~ [L_m, \bar L_n] =0 \cr
[L_n, G_{s,r}] &=&\frac{1}{2} (n-2r) \, G_{s,n+r}, ~~ [\bar L_n, G_{s,r} ] = \frac{1}{2} (n-2s) \, G_{n+s,r}, \cr
 && ~~  [G_{s,r}, G_{s',r'}] = 2 \, \epsilon_{rr'} \, \bar L_{s+s'} + 2 \, \epsilon_{ss'} \, L_{r+r'}
\eea
for $m,n \cdots = 0, \pm 1$, $r,s, \cdots = \pm 1/2$. Here one may think of $L_n$ and $\bar L_n$ to generate either the $\mathfrak{so}(2,2) = \mathfrak{sl}(2, {\mathbb R}) \oplus  \mathfrak{sl}(2, {\mathbb R})$ subalgebra or $\mathfrak{so}(1,3) = \mathfrak{sl}(2, {\mathbb C}) \oplus \mathfrak{sl}(2, {\mathbb C})^c$ subalgebra of $\mathfrak{so}(2,3)$. With the identification: $\bar L_a = J_{a,0}$ and noticing that ${f_{ab}}^c = (a-b) \delta^c_{a+b}$, the first two lines have the same form as those of our algebra (now restricted to $m,n \cdots = 0, \pm 1$, $r,s, \cdots = \pm 1/2$). To compare the $[G_{s,r}, G_{s',r'}]$ let us first take the large-$\kappa$ (the ``classical'') limit in (\ref{GG-comm}). In this limit, from (\ref{final-result-one}, \ref{final-result-two}) we have\footnote{It is curious that in this classical limit the Virasoro central charge approaches the same value as in \cite{Avery:2013dja, Apolo:2014tua, Poojary:2014ifa}, namely $c= -6\kappa$.}  
\bea
 \alpha \rightarrow-\frac{1}{2} \gamma \kappa^2, ~~ 	\beta \rightarrow \frac{1}{2} \gamma \kappa, ~~~ \delta \rightarrow - \gamma \kappa\, ,~~ c\rightarrow -6\,\kappa ~ .
\eea
Further in this limit by choosing $\gamma\rightarrow-\frac{2}{\kappa}$ we arrive at
\bea
\alpha \rightarrow \kappa, ~~  \beta\rightarrow -1, ~~ 	\gamma\rightarrow-\frac{2}{\kappa}\,, ~~ \delta \rightarrow 2\, ,~~ c\rightarrow -6\,\kappa \, . \label{larhe c limit}
\eea
Finally noticing that in (\ref{GG-comm}) the first term proportional to $\alpha$ drops out when restricted to $r,r' = \pm 1/2$. The term proportional to $\delta$, for $r,r' = \pm 1/2$ can be rewritten, using $(r-r') = - \epsilon_{rr'}$ and $(\lambda^a)_{ss'} = \frac{1}{2} \delta^a_{s+s'} $, as 
\bea
\delta \, (r-r') \, J_{a, r+r'} \, {(\lambda^a)}_{ss'} \rightarrow - \epsilon_{rr'} \bar L_{s+s'} ~ .
\eea
Therefore we conclude that the commutator (\ref{GG-comm}) in the large-$\kappa$ limit and restricted to the global modes ($r,r',s,s'\cdots = \pm 1/2$) reads
$$[G_{s,r}, G_{s',r'}] = 2 \, \epsilon_{rr'} \, \bar L_{s+s'} + 2 \, \epsilon_{ss'} \, L_{r+r'} + {\cal O}(1/\kappa).$$
Thus the full algebra (\ref{RESULT}) is therefore an extension (and a deformation around large-$\kappa$) of $\mathfrak{so}(2,3)$ algebra. This completes our analysis.

\subsection*{Comparison with the $\mathcal{W}$-algebra of AdS$_4$ gravity}
In the large $\kappa$-limit, the coefficients $\{\alpha,\beta,\delta, c\}$ take the value as in \eqref{larhe c limit}. In this limit the OPEs become,
\begin{align}
	G_s(z)\,G_{s'}(w)&= \frac{2\kappa\, \epsilon_{ss'}}{(z-w)^3}- \frac{\epsilon_{ss'}\,T(w)}{(z-w)}- \frac{\epsilon_{ss'}\,\frac{2}{\kappa}\, J^2(w)}{(z-w)} \nonumber \\&+ 2\,(\lambda)^a_{ss'}\,\bigg[\frac{\partial J_a (w)}{(z-w)}+ \frac{2\,J_a(w)}{(z-w)^2}\bigg]\\
	T(z)\,T(w)&= -\frac{3\,\kappa}{(z-w)^4}+ \frac{2 T(w)}{(z-w)^2}+ \frac{\partial T(w)}{(z-w)}
\end{align} 
The rest of the other OPEs in this limit do not change. 
Then for the value of $\kappa=2$ and $c=-12$, these OPEs coincide with the  OPEs in \eqref{OPE_relation} (with the following redefinition $G_s \rightarrow \frac{\sqrt{G}}{l}G_s$) that were derived using the postulated line integral charge \eqref{charge}. Therefore the currents that form the representation of chiral $\Lambda$-$\mathfrak{bms}_4$ in AdS$_4$ gravity obey a $\mathcal{W}$-algebra which is the semi-classical limit of $\mathcal{W}(2;\left(3/2\right)^2,1^3)$ that we derived in this section.
\section{Conclusions}
\label{sec7}
In this paper, we generalised the chiral boundary conditions of $\mathbb{R}^{1,3}$ gravity in \cite{Gupta:2021cwo} to AdS$_4$ gravity and show the existence of infinite dimensional algebra as the asymptotic symmetry algebra of chiral locally AdS$_4$ solutions. This algebra is a chiral version of $\Lambda$-$\mathfrak{bms}_4$ algebra \cite{Compere:2019bua,Compere:2020lrt} which we denote as chiral $\Lambda$-$\mathfrak{bms}_4$ algebra. These chiral locally AdS$_4$ solutions obtained after imposing a well-defined variational principle are parameterised by six holomorphic functions  $(T(z), J_a(z), G_s(z))$ for $a = 0, \pm 1$ and $s = \pm 1/2$. We postulated line integral charge from the bulk AdS$_4$ gravity that induces the variations of these holomorphic currents. The charge algebra obeyed a non-linear symmetry algebra and is the semi-classical limit of a $\mathcal{W}$-algebra. To validate our findings, we derive this $\mathcal{W}$-algebra from $2d$ chiral CFT techniques. We show that there exists a one-parameter family of chiral ${\cal W}$-algebras generated by six chiral operators $(T(z), J_a(z), G_s(z))$ for $a = 0, \pm 1$ and $s = \pm 1/2$ with dimensions $(2, 1, 3/2)$ which in semi-classical limit matches with the $\mathcal{W}$-algebra derived from AdS$_4$ gravity. This algebra is isomorphic to one of the algebras derived in \cite{Romans:1990ta} where the current algebra is $\mathfrak{sp}(2)$. In the $\kappa \rightarrow \infty$ limit it admits $\mathfrak{so}(2,3)$ as a subalgebra. Just as  $\mathfrak{so}(2,3)$ admits a contraction to $\mathfrak{iso}(1,3)$, the ${\cal W}$-algebra (\ref{RESULT}) also admits a contraction ($\gamma \rightarrow 0$ for finite $\kappa$) to chiral $\mathfrak{bms}_4$ algebra that appeared in the studies of graviton soft-theorems \cite{Banerjee:2020zlg, Banerjee:2021dlm}.

Using the nomenclature of \cite{Bouwknegt:1992wg} we refer to this new algebra (\ref{RESULT}) as ${\cal W}(2;( 3/2)^2, 1^3)$. Even though ${\cal W}(2;( 3/2)^2, 1^3)$ has a chiral primary $G_s(z)$ of half-integer dimension, it is not a superconformal algebra such as the one of Knizhnik \cite{Knizhnik:1986wc}. The reason for this is that even though the current $G_s$ has $h=3/2$ it still has an integer spin defined by $L_0 - J_0$. The algebra (\ref{RESULT}) is more akin to the Bershadsky-Polyakov algebra ${\cal W}_3^{(2)}$ -- where by considering a twisted stress tensor one can make $G_s(z)$ to appear bosonic (integral dimension) \cite{Bershadsky:1990bg, Polyakov:1989dm}.

The existence of line integral charges \eqref{charge} from bulk AdS$_4$ gravity needs to be understood better. As mentioned previously, an induced gravity action at the boundary can lead to such line integral charges. For locally AdS$_4$ solution as shown by Skenderis and Solodukhin in \cite{Skenderis:1999nb},  the Weyl and diffeomorphism invariant boundary effective action is zero therefore one needs to revisit the arguments of \cite{Skenderis:1999nb} and provide a prescription to calculate such an effective action for locally AdS$_4$ configurations. This approach is similar to \cite{Nguyen:2020hot}, where the 2d effective action for Goldstone mode associated with the spontaneous symmetry breaking of  superrotation symmetry  of 4d locally flat solutions was derived using Hamiltonian reduction of $\mathfrak{sl}(2,\mathbb C)$ CS theory. This 2d CFT gives rise to line integral charges for such modes. It was further shown in \cite{Nguyen:2022zgs} that this CFT corresponds to the unobservable gauge sector within the context of flat space holography. Since the flat space limit of locally AdS$_4$ solutions derived in this paper include these Goldstone modes, it will be interesting to understand the role played by  these solutions in the context of AdS holography. Recently in \cite{Ciambelli:2024kre} by considering the radiation in AdS spacetime, the authors extracted celestial currents from the flat space limit of such AdS solutions. It will be interesting to show the existence of our $\mathcal W$-algebra in their setup.

In this paper, we looked at the locally AdS$_4$ solutions as we wanted to compare the flat limit of it to the locally flat solution of \cite{Gupta:2021cwo}. As mentioned before, locally AdS$_4$ condition implies the vanishing of Cotton tensor of $3d$ induced boundary metric \eqref{bdrymetric}. After imposing  $v_a=0$ to satisfy variational principle and considering  $2d$ boundary metric in Polyakov gauge \eqref{co-dimension-two}, the vanishing of Cotton tensor leads to the form \eqref{co-dimension two 2} of the only non-zero function of the boundary metric \eqref{bdrymetric}. However, one can definitely consider solutions that are not locally AdS$_4$, but continue to share our boundary conditions with conformally flat boundary metrics. One expects these spacetimes to form a subset of asymptotically locally AdS$_4$ solutions with leaky boundary conditions considered in \cite{Compere:2019bua, Compere:2020lrt} for which the asymptotic symmetry algebra was shown to be a Lie algebroid $\Lambda$-$\mathfrak{bms}_4$. The choice of boundary metric that we consider allows us to solve the constraint equations of $\Lambda$-$\mathfrak{bms}_4$ vector fields in \cite{Compere:2019bua} in terms of holomorphic functions \eqref{bdry diff}. As a result, our chiral $\Lambda$-$\mathfrak{bms}_4$ symmetries given in terms of  \eqref{bdry diff} are a subset of asymptotic vector fields that form the $\Lambda$-$\mathfrak{bms}_4$ algebra in \cite{Compere:2019bua}.  To summarize, one can continue to impose our chiral boundary conditions on more general asymptotically (not necessarily locally) AdS$_4$ configurations, and one expects these configurations to be a subset of configurations allowed by $\Lambda$-$\mathfrak{bms}_4$ boundary conditions of \cite{Compere:2019bua}.
 
Another chiral $\mathcal{W}$-algebra extension of $\mathfrak{so}(2,3)$ in the semi-classical limit was obtained in \cite{Fuentealba:2020zkf} from the asymptotic symmetry analysis of $3d$ conformal gravity. This algebra was called conformal $\mathfrak{bms}_3$ or $W_{(2,2,2,1)}$. The derivation of the full quantum version of this algebra for finite $c$ and $\kappa$ and other possible chiral $\mathcal{W}$-algebra extensions of $\mathfrak{so}(2,3)$ and their derivation from AdS$_4$ gravity will be explored in our future work.

\appendix
\bibliographystyle{utphys}

\begin{thebibliography}{10}
\bibitem{Zamolodchikov:1985wn}
A.~B.~Zamolodchikov,
``Infinite Additional Symmetries in Two-Dimensional Conformal Quantum Field Theory,''
Theor. Math. Phys. \textbf{65} (1985), 1205-1213
doi:10.1007/BF01036128

\bibitem{Bouwknegt:1992wg}
P.~Bouwknegt and K.~Schoutens,
``W symmetry in conformal field theory,''
Phys. Rept. \textbf{223} (1993), 183-276
doi:10.1016/0370-1573(93)90111-P
[arXiv:hep-th/9210010 [hep-th]].

\bibitem{Henneaux:2010xg}
M.~Henneaux and S.~J.~Rey,
``Nonlinear $W_{infinity}$ as Asymptotic Symmetry of Three-Dimensional Higher Spin Anti-de Sitter Gravity,''
JHEP \textbf{12} (2010), 007
doi:10.1007/JHEP12(2010)007
[arXiv:1008.4579 [hep-th]].

\bibitem{Campoleoni:2010zq}
A.~Campoleoni, S.~Fredenhagen, S.~Pfenninger and S.~Theisen,
JHEP \textbf{11} (2010), 007
doi:10.1007/JHEP11(2010)007
[arXiv:1008.4744 [hep-th]].

\bibitem{Banerjee:2020zlg}
S.~Banerjee, S.~Ghosh and P.~Paul,
``MHV graviton scattering amplitudes and current algebra on the celestial sphere,''
JHEP \textbf{02} (2021), 176
doi:10.1007/JHEP02(2021)176
[arXiv:2008.04330 [hep-th]].

\bibitem{Banerjee:2021dlm}
S.~Banerjee, S.~Ghosh and P.~Paul,
``(Chiral) Virasoro invariance of the tree-level MHV graviton scattering amplitudes,''
JHEP \textbf{09} (2022), 236
doi:10.1007/JHEP09(2022)236
[arXiv:2108.04262 [hep-th]].

\bibitem{Gupta:2021cwo}
N.~Gupta, P.~Paul and N.~V.~Suryanarayana,
``An $\widehat{sl_2}$ Symmetry of ${\mathbb R}^{1,3}$ Gravity,''
[arXiv:2109.06857 [hep-th]].


\bibitem{Mishra:2017zan}
R.~K.~Mishra and R.~Sundrum,
JHEP \textbf{01} (2018), 014
doi:10.1007/JHEP01(2018)014
[arXiv:1706.09080 [hep-th]].


\bibitem{Lowe:2020qan}
D.~A.~Lowe and D.~M.~Ramirez,
JHEP \textbf{01} (2021), 075
doi:10.1007/JHEP01(2021)075
[arXiv:2007.02851 [hep-th]].

\bibitem{Taylor:2023ajd}
T.~R.~Taylor and B.~Zhu,
Phys. Rev. Lett. \textbf{132}, no.22, 221602 (2024)
doi:10.1103/PhysRevLett.132.221602
[arXiv:2312.00876 [hep-th]].

\bibitem{Bittleston:2024rqe}
R.~Bittleston, G.~Bogna, S.~Heuveline, A.~Kmec, L.~Mason and D.~Skinner,
JHEP \textbf{07}, 010 (2024)
doi:10.1007/JHEP07(2024)010
[arXiv:2403.18011 [hep-th]].

\bibitem{Compere:2019bua}
G.~Comp\`ere, A.~Fiorucci and R.~Ruzziconi,
``The $\Lambda$-BMS$_4$ group of dS$_4$ and new boundary conditions for AdS$_4$,''
Class. Quant. Grav. \textbf{36} (2019) no.19, 195017


\bibitem{Compere:2020lrt}
G.~Comp\`ere, A.~Fiorucci and R.~Ruzziconi,
``The $\Lambda$-BMS$_4$ charge algebra,''
JHEP \textbf{10} (2020), 205

\bibitem{Gupta:2020dtl}
N.~Gupta and N.~V.~Suryanarayana,
``Constructing Carrollian CFTs,''
JHEP \textbf{03}, 194 (2021)
doi:10.1007/JHEP03(2021)194
[arXiv:2001.03056 [hep-th]].

\bibitem{Compere:2008us}
G.~Compere and D.~Marolf,
``Setting the boundary free in AdS/CFT,''
Class. Quant. Grav. \textbf{25} (2008), 195014
doi:10.1088/0264-9381/25/19/195014
[arXiv:0805.1902 [hep-th]].



\bibitem{Skenderis:1999nb}
K.~Skenderis and S.~N.~Solodukhin,
``Quantum effective action from the AdS / CFT correspondence,''
Phys. Lett. B \textbf{472} (2000), 316-322
doi:10.1016/S0370-2693(99)01467-7
[arXiv:hep-th/9910023 [hep-th]].
\bibitem{Marolf:2012vvz}
D.~Marolf, W.~Kelly and S.~Fischetti,
doi:10.1007/978-3-642-41992-8\_19
[arXiv:1211.6347 [gr-qc]].
\bibitem{Poole:2018koa}
A.~Poole, K.~Skenderis and M.~Taylor,
Class. Quant. Grav. \textbf{36} (2019) no.9, 095005
doi:10.1088/1361-6382/ab117c
[arXiv:1812.05369 [hep-th]].

\bibitem{Romans:1990ta}
L.~J.~Romans,
``Quasisuperconformal algebras in two-dimensions and Hamiltonian reduction,''
Nucl. Phys. B \textbf{357} (1991), 549-564


\bibitem{Blumenhagen:1990jv}
R.~Blumenhagen, M.~Flohr, A.~Kliem, W.~Nahm, A.~Recknagel and R.~Varnhagen,
``W algebras with two and three generators,''
Nucl. Phys. B \textbf{361} (1991), 255-289
doi:10.1016/0550-3213(91)90624-7
\bibitem{Belavin:1984vu}
A.~A.~Belavin, A.~M.~Polyakov and A.~B.~Zamolodchikov,
``Infinite Conformal Symmetry in Two-Dimensional Quantum Field Theory,''
Nucl. Phys. B \textbf{241} (1984), 333-380
doi:10.1016/0550-3213(84)90052-X


\bibitem{DiFrancesco:1997nk}
P.~Di Francesco, P.~Mathieu and D.~Senechal,
``Conformal Field Theory,''
Springer-Verlag, 1997,
ISBN 978-0-387-94785-3, 978-1-4612-7475-9
doi:10.1007/978-1-4612-2256-9

\bibitem{Blumenhagen:2009zz}
R.~Blumenhagen and E.~Plauschinn,
``Introduction to conformal field theory: with applications to String theory,''
Lect. Notes Phys. \textbf{779} (2009), 1-256
doi:10.1007/978-3-642-00450-6

\bibitem{Thielemans:1994er}
K.~Thielemans,
``An Algorithmic approach to operator product expansions, W algebras and W strings,''
[arXiv:hep-th/9506159 [hep-th]].

\bibitem{Barnich:2011ty}
G.~Barnich and P.~H.~Lambert,
``A Note on the Newman-Unti group and the BMS charge algebra in terms of Newman-Penrose coefficients,''
Adv. Math. Phys. \textbf{2012} (2012), 197385
doi:10.1155/2012/197385
[arXiv:1102.0589 [gr-qc]].


\bibitem{Balasubramanian:1999re}
V.~Balasubramanian and P.~Kraus,
Commun. Math. Phys. \textbf{208}, 413-428 (1999)
doi:10.1007/s002200050764
[arXiv:hep-th/9902121 [hep-th]].

\bibitem{Brown:1992br}
J.~D.~Brown and J.~W.~York, Jr.,
Phys. Rev. D \textbf{47}, 1407-1419 (1993)
doi:10.1103/PhysRevD.47.1407
[arXiv:gr-qc/9209012 [gr-qc]].

\bibitem{Bhattacharyya:2007vjd}
S.~Bhattacharyya, V.~E.~Hubeny, S.~Minwalla and M.~Rangamani,
JHEP \textbf{02}, 045 (2008)
doi:10.1088/1126-6708/2008/02/045
[arXiv:0712.2456 [hep-th]].



\bibitem{Compere:2018ylh}
G.~Comp\`ere, A.~Fiorucci and R.~Ruzziconi,
``Superboost transitions, refraction memory and super-Lorentz charge algebra,''
JHEP \textbf{11} (2018), 200
[erratum: JHEP \textbf{04} (2020), 172]
doi:10.1007/JHEP11(2018)200
[arXiv:1810.00377 [hep-th]].
\bibitem{Avery:2013dja}
S.~G.~Avery, R.~R.~Poojary and N.~V.~Suryanarayana,
``An sl(2,$\mathbb{R}$) current algebra from $AdS_3$ gravity,''
JHEP \textbf{01} (2014), 144
doi:10.1007/JHEP01(2014)144
[arXiv:1304.4252 [hep-th]].
\bibitem{Barnich:2010eb}
G.~Barnich and C.~Troessaert,
``Aspects of the BMS/CFT correspondence,''
JHEP \textbf{05} (2010), 062
doi:10.1007/JHEP05(2010)062
[arXiv:1001.1541 [hep-th]].

\bibitem{Wald:1999wa}
R.~M.~Wald and A.~Zoupas,
Phys. Rev. D \textbf{61} (2000), 084027
doi:10.1103/PhysRevD.61.084027
[arXiv:gr-qc/9911095 [gr-qc]].

\bibitem{Iyer:1994ys}
V.~Iyer and R.~M.~Wald,
Phys. Rev. D \textbf{50} (1994), 846-864
doi:10.1103/PhysRevD.50.846
[arXiv:gr-qc/9403028 [gr-qc]].
\bibitem{Lee:1990nz}
J.~Lee and R.~M.~Wald,
J. Math. Phys. \textbf{31} (1990), 725-743
doi:10.1063/1.528801
\bibitem{Barnich:2001jy}
G.~Barnich and F.~Brandt,
``Covariant theory of asymptotic symmetries, conservation laws and central charges,''
Nucl. Phys. B \textbf{633} (2002), 3-82
doi:10.1016/S0550-3213(02)00251-1
[arXiv:hep-th/0111246 [hep-th]].
\bibitem{Barnich:2007bf}
G.~Barnich and G.~Compere,
``Surface charge algebra in gauge theories and thermodynamic integrability,''
J. Math. Phys. \textbf{49} (2008), 042901
doi:10.1063/1.2889721
[arXiv:0708.2378 [gr-qc]].
\bibitem{Barnich:2000zw}
G.~Barnich, F.~Brandt and M.~Henneaux,
``Local BRST cohomology in gauge theories,''
Phys. Rept. \textbf{338} (2000), 439-569
doi:10.1016/S0370-1573(00)00049-1
[arXiv:hep-th/0002245 [hep-th]].
\bibitem{Compere:2016jwb}
G.~Comp\`ere and J.~Long,
JHEP \textbf{07}, 137 (2016)
doi:10.1007/JHEP07(2016)137
[arXiv:1601.04958 [hep-th]].


\bibitem{Bowcock:1990ku}
P.~Bowcock,
``Quasi-primary Fields and Associativity of Chiral Algebras,''
Nucl. Phys. B \textbf{356} (1991), 367-386
doi:10.1016/0550-3213(91)90314-N

\bibitem{Polyakov:1987zb}
A.~M.~Polyakov,
``Quantum Gravity in Two-Dimensions,''
Mod. Phys. Lett. A \textbf{2} (1987), 893
doi:10.1142/S0217732387001130

\bibitem{Apolo:2014tua}
L.~Apolo and M.~Porrati,
``Free boundary conditions and the AdS$_3$/CFT$_2$ correspondence,''
JHEP \textbf{03} (2014), 116
doi:10.1007/JHEP03(2014)116
[arXiv:1401.1197 [hep-th]].

\bibitem{Poojary:2014ifa}
R.~R.~Poojary and N.~V.~Suryanarayana,
``Holographic chiral induced W-gravities,''
JHEP \textbf{10} (2015), 168
doi:10.1007/JHEP10(2015)168
[arXiv:1412.2510 [hep-th]].

\bibitem{Knizhnik:1986wc}
V.~G.~Knizhnik,
``Superconformal Algebras in Two-dimensions,''
Theor. Math. Phys. \textbf{66} (1986), 68-72
doi:10.1007/BF01028940

\bibitem{Bershadsky:1990bg}
M.~Bershadsky,
``Conformal field theories via Hamiltonian reduction,''
Commun. Math. Phys. \textbf{139} (1991), 71-82
doi:10.1007/BF02102729
\bibitem{Polyakov:1989dm}
A.~M.~Polyakov,
``Gauge Transformations and Diffeomorphisms,''
Int. J. Mod. Phys. A \textbf{5} (1990), 833
doi:10.1142/S0217751X90000386

\bibitem{Nguyen:2020hot}
K.~Nguyen and J.~Salzer,
JHEP \textbf{02}, 108 (2021)
doi:10.1007/JHEP02(2021)108
[arXiv:2008.03321 [hep-th]].

\bibitem{Nguyen:2022zgs}
K.~Nguyen,
PoS \textbf{CORFU2021}, 133 (2022)
doi:10.22323/1.406.0133
[arXiv:2201.09640 [hep-th]].
\bibitem{Ciambelli:2024kre}
L.~Ciambelli, S.~Pasterski and E.~Tabor,
``Radiation in Holography,''
[arXiv:2404.02146 [hep-th]].

\bibitem{Fuentealba:2020zkf}
O.~Fuentealba, H.~A.~Gonz\'alez, A.~P\'erez, D.~Tempo and R.~Troncoso,
Phys. Rev. Lett. \textbf{126} (2021) no.9, 091602
doi:10.1103/PhysRevLett.126.091602
[arXiv:2011.08197 [hep-th]].




\end{thebibliography}
\providecommand{\href}[2]{#2}\begingroup\raggedright\endgroup 
\end{document}